\documentclass[fleqn,usenatbib]{mnras}

\usepackage{newtxtext,newtxmath}
\usepackage{soul,xcolor}

\usepackage[T1]{fontenc}
\usepackage{ae,aecompl}

\usepackage{graphicx}	
\usepackage{amsmath}	
\usepackage{amssymb}	
\usepackage{xcolor}     
\usepackage{xspace}

\newcommand{\sfr}{M${_\odot}$ yr$^{-1}$}	
\newcommand{\msun}{M${_\odot}$}	
\newcommand{\lsun}{L${_\odot}$}	
\newcommand{\jup}{${\rm J_{\rm up}}$}	
\newcommand{\kms}{${\rm km~s^{-1}}$}

\title[High-J CO lines in $z\sim 6$ quasars]{Constraints on high-J CO emission lines in $z\sim 6$ quasars}

\author[Stefano Carniani et al.]{S.~Carniani$^{1}$\thanks{E-mail: stefano.carniani@sns.it},
S.~Gallerani$^{1}$,
L.~Vallini$^{2,3}$,
A.~Pallottini$^{4,1}$,
M.~Tazzari$^{5}$,
A.~Ferrara$^{1}$,
\newauthor
R.~Maiolino$^{6,7}$, 
C.~Cicone$^{8}$,
C.~Feruglio$^{9}$, 
R.~Neri$^{10}$,
V.~D'Odorico$^{11,1}$,
R.~Wang$^{12}$,
 J.~Li$^{13}$ \\
$^{1}$ Scuola Normale Superiore, Piazza dei Cavalieri 7, I-56126 Pisa, Italy \\
$^{2}$ Leiden Observatory, P.O. Box 9513 NL-2300 RA, (NL) \\
$^{3}$ Nordita, KTH Royal Institute of Technology and Stockholm University, Roslagstullsbacken 23, SE-10691 Stockholm, Sweden \\
$^{4}$ Centro Fermi, Museo Storico della Fisica e Centro Studi e Ricerche `Enrico Fermi', Piazza del Viminale 1, I-00184 Roma, Italy \\
$^{5}$ Institute of Astronomy, University of Cambridge, Madingley Road, Cambridge CB3 0HA, UK \\
$^{6}$ Kavli Institute for Cosmology, University of Cambridge, Madingley Road, Cambridge CB3 0HA, UK
 \\
$^{7}$ Cavendish Laboratory, University of Cambridge, 19 J. J. Thomson Ave., Cambridge CB3 0HE, UK\\
$^{8}$ INAF - Osservatorio Astronomico di Brera, Via Brera 28, 20121 Milano, Italy\\
$^{9}$ INAF Osservatorio Astronomico di Trieste, Via G. Tiepolo 11, Trieste, Italy \\
$^{10}$ Institute de Radioastronomie Millimetrique, St. Martin d'Heres, F-38406, France \\
$^{11}$ INAF Osservatorio Astronomico di Trieste, via G. Tiepolo 11, Trieste, Italy \\
$^{12}$ Kavli Institute of Astronomy and Astrophysics at Peking University, No.5 Yiheyuan Road, Haidian District, Beijing, 100871, China\\
$^{13}$ Department of Astronomy, School of Physics, Peking University, Beijing 100871, China
}
\date{Accepted XXX. Received YYY; in original form ZZZ}

\pubyear{2019}
\begin{document}

\setstcolor{red}

\label{firstpage}
\pagerange{\pageref{firstpage}--\pageref{lastpage}}
\maketitle

\begin{abstract}
We present Atacama Large Millimiter/submillimiter Array (ALMA) observations of eight highly excited CO (\jup~$>8$) lines and continuum emission in two $z\sim6$ quasars: SDSS J231038.88+185519.7 (hereafter J2310), for which CO(8-7), CO(9-8), and CO(17-16) lines have been observed, and ULAS J131911.29+095951.4 (J1319), observed in the CO(14-13), CO(17-16) and CO(19-18) lines. 
The continuum  emission of both quasars arises from a compact region ($< 0.9$ kpc). By assuming a modified black-body law, we estimate  dust masses of Log$(M_{\rm dust}/M_{\odot})=8.75\pm0.07$ and Log$(M_{\rm dust}/M_{\odot})=8.8\pm0.2$ and  dust temperatures of $T_{\rm dust}=76\pm3~{\rm K}$  and $T_{\rm dust}=66^{+15}_{-10}~{\rm K}$, respectively for J2310 and J1319. 
Only CO(8-7) and CO(9-8) in J2310 are detected, while $3\sigma$ upper limits on luminosities are reported for the other lines of both quasars.
The CO line luminosities and upper limits measured in J2310 and J1319 are consistent with those observed in local AGN and starburst galaxies, and other $z\sim 6$ quasars, except for SDSS J1148+5251 (J1148), the only quasar at $z=6.4$ with a previous CO(17-16) line detection.
By computing the CO SLEDs normalised to the CO(6-5) line and FIR luminosities for J2310, J1319, and J1149, we conclude that  different gas heating mechanisms (X-ray radiation and/or shocks)  may explain the different CO luminosities observed in these $z\sim6$ quasar.
Future \jup~$>8$ CO observations will be crucial to understand the processes responsible for molecular gas excitation in luminous high-$z$ quasars.
\end{abstract}

\begin{keywords}
quasars: individual: SDSS J231038.88+185519.7 - quasars: individual: ULAS J131911.29+095951.4 - galaxies: high-redshift - galaxies: active - galaxies: ISM
\end{keywords}



\section{Introduction}
\label{sec:introduction}
The presence of early super massive black holes (SMBH) represents a challenging problem in modern cosmology. In the last decades, more than 200 quasars have been discovered at $z\sim6$ and beyond \citep[e.g.][]{ Banados:2016, Banados:2018, Jiang:2016, Matsuoka:2016, Matsuoka:2018, Mazzucchelli:2017}; for several of them, the mass of the hosted BH has been measured and found to be $M_{\rm BH} \sim (0.02-1.1)\times10^{10} M_{\odot}$ \citep{Kurk:2007, Jiang:2007, Willott:2010, De-Rosa:2011, Wu:2015, Feruglio:2018}. The presence of such supermassive BHs when the Universe was less than 1 Gyr old, is still an open problem \citep{Li:2007, Narayanan:2008, Volonteri:2010, Di-Matteo:2012, Valiante:2017}, deeply connected both with the galaxy-BH co-evolution \citep[e.g.][]{Wang:2010, Lamastra:2010, Volonteri:2011, Valiante:2014} and the contribution of quasars to the cosmic reionization process \citep{Volonteri:2009, Giallongo:2015a, Madau:2015a, Manti:2017, Qin:2017, Parsa:2018, Mitra:2018, Kulkarni:2018}.

In the last years, the advent of millimetre and submillimeter interferometers has given the possibility of studying the physical properties of the interstellar medium (ISM) in $z\sim 6$ galaxies hosting SMBHs. In fact, several important tracers of the ISM physical and chemical state, such as the [\ion{C}{ii}] line at 158$\mu$m, CO rotational transitions, and dust continuum emission, are redshifted in the millimetre bands at high redshift and are observable from ground-based facilities \citep[see][for a review on this topic]{Gallerani:2017a}. The Atacama Large Millimiter/submillimiter Array (ALMA) is currently the most powerful interferometer for observing rest-frame far-infrared (FIR) emission in the distant Universe, as highlighted in  several recent works showing ALMA capabilities on investigating ISM properties in $z>4$ quasars \citep{Gallerani:2012,Carniani:2013, Venemans:2017,Venemans:2017a, Decarli:2017, Decarli:2018, Bischetti:2018, Feruglio:2018}. 

In this work we focus on CO, the most abundant molecule after H$_2$ \citep{Carilli:2013}. CO has a finite dipole moment that allows transitions between energy levels with small energy gaps. Examining  the CO spectral line energy distribution (SLED), which is the relative luminosity (or intensity)  of CO lines as a function of rotational transitions \jup,   provides us the opportunity to probe the excitation conditions of molecular gas in galaxies. This kind of observations will not be possible with the fainter quadrupole transitions of H$_2$ until the advent of SPace IR telescope for Cosmology and Astrophysics  \citep[SPICA;][]{Spinoglio:2017, Egami:2018}

The strength of low-J and mid-J transitions (\jup~$<6$) is mainly driven by physical properties such as gas density and temperature \citep[e.g.][]{Obreschkow:2009,Mashian:2015}.
At high redshift ($z\sim 6$) the shape of the CO SLED also depends on the cosmic microwave background (CMB) radiation \citep{da-Cunha:2013}, since the CMB temperature ($\rm T_{\rm CMB}\sim 20$ K) becomes comparable to that of the cold/molecular gas ($\sim 10-50$ K). The high $\rm T_{CMB}$  increases the gas excitation conditions and boosts the CO line intensities. At the same time, the CMB background against which the line is measured increases too. This twofold effect shifts the peak of the  CO SLED at higher transitions up to \jup~$\sim6-7$ \citep{Tunnard:2016,Vallini:2018}. 
 Mid-J CO observations  can  be therefore used in the early Universe  for investigating the molecular gas reservoir and ISM properties in both star-forming  \citep[e.g.][]{Genzel:2015, Aravena:2016} and quasar host galaxies \cite[e.g.][]{Kakkad:2017, Carniani:2017,Venemans:2017, Venemans:2017a,Brusa:2018}.

High-J (\jup~$\geq 7$) CO lines are only emitted from states with temperatures 150~-~7000 K above the ground and have critical densities of $10^5-10^8~{\rm cm}^{-3}$.
These lines trace molecular warm and dense gas and are often over-luminous in extreme environments, such as luminous AGN \citep[e.g.][]{Meijerink:2007,Schleicher:2010}, extreme starbursts \citep[SFR $ > 1000$ \sfr, e.g.][]{Narayanan:2008}, regions shocked by merging or outflows mechanisms \citep{panuzzo:2010,Hailey-Dunsheath:2012, Richings:2018}.
Identifying the dominant mechanism for molecular gas excitation is crucial for a proper interpretation of high-J CO line observations, and thus for a deeper understanding of the ISM properties. 

In the local Universe high-J CO lines have been detected by using the Photodetector Array Camera and Spectrometer \citep[PACS,][]{Poglitsch:2010} on board of the Herschel Space Observatory \citep{Pilbratt:2010}.  \cite{Mashian:2015} reported the high-J CO SLED ($14 \leq$~\jup~$ \leq 50$) of 5 starburst galaxies, 5 AGN, 22 ULIRGs and 2 interacting systems. They found that the extreme diversity in CO emission makes multiple lines essential to constrain the gas properties.

At high-$z$, \jup~$>7$ CO transitions have been observed only in quasars and extreme starburst galaxies \citep{Weis:2007,Riechers:2013, Gallerani:2014,Venemans:2017a}. In particular \cite{Gallerani:2014} detected with the Plateau de Bure interferometer (PdBI) an exceptionally strong CO(17-16) line in the $z = 6.4$ quasar SDSS J114816.64+525150.3 (hereafter J1148). By combining previous CO observations  \citep{Bertoldi:2003a, Walter:2003, Riechers:2009} with the detection of the CO(17-16), and by comparing the observed CO SLED with Photo-Dissociation Regions (PDR) and X-ray Dominated Region (XDR) models \citep{Meijerink:2005,Meijerink:2007}, the authors found that while PDR models can fairly reproduce the observed CO SLED for \jup~$ < 7$, the CO(17-16) line can only be explained through the presence of a substantial X-ray radiation field. Indeed, strong X-ray emission ($L_X\sim 10^{45}\rm erg~s^{-1}$) was recently detected in this source \citep{Gallerani:2017}, supporting the idea that high-J CO transitions may be used to infer the presence of X-ray faint or obscured SMBH progenitors in galaxies at $z>6$. Since J1148 is the unique source where high-J CO lines have been detected so far at high redshift, these results motivated further observations of highly excited CO lines in $z\sim 6$ quasars. 

In this work we present ALMA observations of six \mbox{high-J} CO lines and continuum emission in the $z\sim 6$ quasars  SDSS J231038.88+185519.7 (hereafter J2310) at  $z=6.00$ and ULAS J131911.29+095951.4 (hereafter J1319) at $z=6.13$. We  re-analyse the FIR emission and CO SLED of  SDSS J1148+5251, and compare the properties of these three quasars.
The paper is organised as follows: the two targets are presented in Sec.~\ref{sec:targets}, while  ALMA observations are described in Sec.~\ref{sec:observations}. In Sec.~\ref{sec:results}, we present continuum and CO emission properties of the three quasars. In Sec.~\ref{sec:sled}, we compare our results with local and high-$z$ observations. We discuss and summarise our findings in Sec.~\ref{sec:discussion} and Sec.~\ref{sec:conclusions}, respectively. We adopt the  cosmological parameters  from \cite{Planck-Collaboration:2015}: H$_0$ = 67.7 km s$^{-1}$ Mpc$^{-1}$, $\Omega_{\rm m}$ = 0.308 and  $\Omega_{\rm \Lambda}$ = 0.70, according to which 1\arcsec\ at $z = 6$ corresponds to a proper distance of 5.84~kpc.

\section{Targets}\label{sec:targets}

\subsection{J2310}
With a  magnitude m$_{1450}=19.30$, J2310 is one of the brightest $z\sim6$ quasar in the Sloan Digital Sky Survey. By using the UV lines of \ion{C}{iv} and \ion{Mg}{ii}, \cite{Jiang:2016} and \cite{Feruglio:2018} estimate a BH mass $M_{\rm BH}=(1.8\pm0.5)\times 10^{9}$ \msun\ that is 2.5\% of the dynamical mass recently  inferred from  high-angular ALMA observations of CO(6-5) line \citep{Feruglio:2018}. 
The estimated dynamical and BH masses place J2310 above the local M$_{\rm dyn} - {\rm M}_{\rm BH}$ relation, similarly to most of the quasars studied at these redshifts \citep{Wang:2013,Decarli:2018}.

Over the last years J2310 has been also studied extensively in the millimetre bands.  FIR line observations of [\ion{C}{ii}]  at 158$\mu$m and  [\ion{O}{iii}]  at 88$\mu$m  have been reported by \cite{Wang:2013} and  \cite{Hashimoto:2018a}, respectively. 

\subsection{J1319}

J1319 was initially discovered in the UKIRT Infrared Deep Sky Survey (UKIDSS) by \cite{Mortlock:2009}, who measured an optical magnitude of m$_{1450}=19.65$ and identified its redshift from the Ly$\alpha$ and \ion{Mg}{ii} emission lines. 
\cite{Shao:2017} estimate a black hole mass $M_{\rm BH}=(2.7\pm0.6)\times 10^9\rm M_{\odot}$ from the \ion{Mg}{II} line,  which contributes 2\% of the dynamical mass of the system. The BH-to-dynamical mass ratio is four times larger than the average $M_{\rm BH}/M_{\rm dyn}$ measured locally \citep{Kormendy:2013}. 

Several millimetre observations have been carried out of this quasar  covering the frequency range between 1.4 GHz and 300 GHz, revealing  a high far-infrared emission ($>10^{13}$ \lsun) from the host galaxy \citep{Wang:2011, Wang:2013}. The quasar has been also detected  in both CO(6-5) and [\ion{C}{ii}] at 158$\mu$m \citep{Wang:2011, Wang:2013}.

\section{Observations}
\label{sec:observations}

In ALMA Cycle 2 project 2013.1.00462.S (P.I. S. Gallerani) we  proposed high-J CO observations of the $z\sim6$ quasars J2310 and J1319 that have been already observed in CO(6-5) \citep{Wang:2013, Feruglio:2018}. We requested ALMA time to observe  the molecular line CO(17-16) at ${\rm \nu_{\rm rest}= 1956.02 ~GHz}$ (${\rm \nu_{\rm obs}=279.31~GHz}$) for J2310, and  three high-J CO transitions, CO(14-13) at ${\rm \nu_{\rm rest}= 1611.79~GHz}$ (${\rm \nu_{\rm obs}=225.96~GHz}$), CO(17-16) (${\rm \nu_{\rm obs}= 274.22~GHz}$), and CO(19-18) at ${\rm \nu_{\rm rest}= 2185.13~GHz}$ (${\rm \nu_{\rm obs}= 306.34~GHz}$),  for J1319. ALMA band 6 and 7  observations were obtained between January and June 2015,  using 34-40 antennas with baselines from 15 m to 700 m. In each observations, one out of the four 1.8 GHz spectral windows was centred at the CO frequency and the other three were used to sample the rest-frame far-infrared (FIR) continuum emission. 
The spectral channel width was set up in time domain mode with a spectral resolution of  31.250 MHz, corresponding to $\sim$20~km~s$^{-1}$ velocity resolution at the line frequencies. 
For each CO observations we spent between 13 and 31 minutes on source, reaching a continuum sensitivity of $\sim42$~$\mu$Jy~beam$^{-1}$ and a spectral sensitivity of 0.35-0.4~mJy~beam$^{-1}$ per spectral channel.

ALMA visibilities have been calibrated using the CASA software \citep{McMullin:2007}. The final continuum images and  datacubes have been generated with the CASA task \texttt{tclean}  using  a natural weighting, which gives the optimum point-source sensitivity in the image plane.  Final products have angular resolutions between 0.6\arcsec\ and 1.5\arcsec\ depending on the datasets (see Table 1). 
The uv-coverage  of the observations results in a largest angular resolution  between (LAS) 1.7\arcsec\ and 5\arcsec\ depending on the observed frequencies.
Continuum images have been obtained from the line-free channels of the four spectral windows that have been also used in the task \texttt{uvcontsub} task to subtract the continuum emission in the uv-plane. The datacubes have been generated from the  continuum-subtracted uv-datasets. 

In addition to the observations of our  programme, we have searched in the ALMA archive for further public datasets. We have thus used the   ALMA observations of the project 2015.1.01265.S (P.I. R. Wang) targeting the CO(8-7) line at ${\rm \nu_{\rm rest}= 921.80~GHz}$ (${\rm \nu_{\rm obs}=131.63~GHz}$) and the CO(9-8) line at ${\rm \nu_{\rm rest}= 1036.91~GHz}$ (${\rm \nu_{\rm obs}=148.06~GHz}$) in J2310. The datasets will be presented in a forthcoming  paper by Shao et al. (2019) and Li et al. (in prep.). In this work, we present the continuum emissions at 132 GHz and 141 GHz and the two CO lines with their relative luminosities. For J1319, we have  benefited   from  continuum emission  at  347 GHz  observed in the ALMA project  2012.1.00391.S (P.I. J. Gracia-Carpio). All retrieved observations have been reduced by following the ALMA pipeline released with the datasets and final image have been generated by adopting a natural weight scheme.

In addition to the ALMA data we have  collected  from the literature all sub/millimetre measurements associated to these two quasars as well as to J1148 (see Table~\ref{tab:cont} and \ref{tab:prop}). 


\section{Results}
\label{sec:results}

\subsection{Continuum emission} 
\label{sec:continuum.emission}
\begin{figure*}
    \centering
    {\huge J2310}\\
	\includegraphics[width=2\columnwidth]{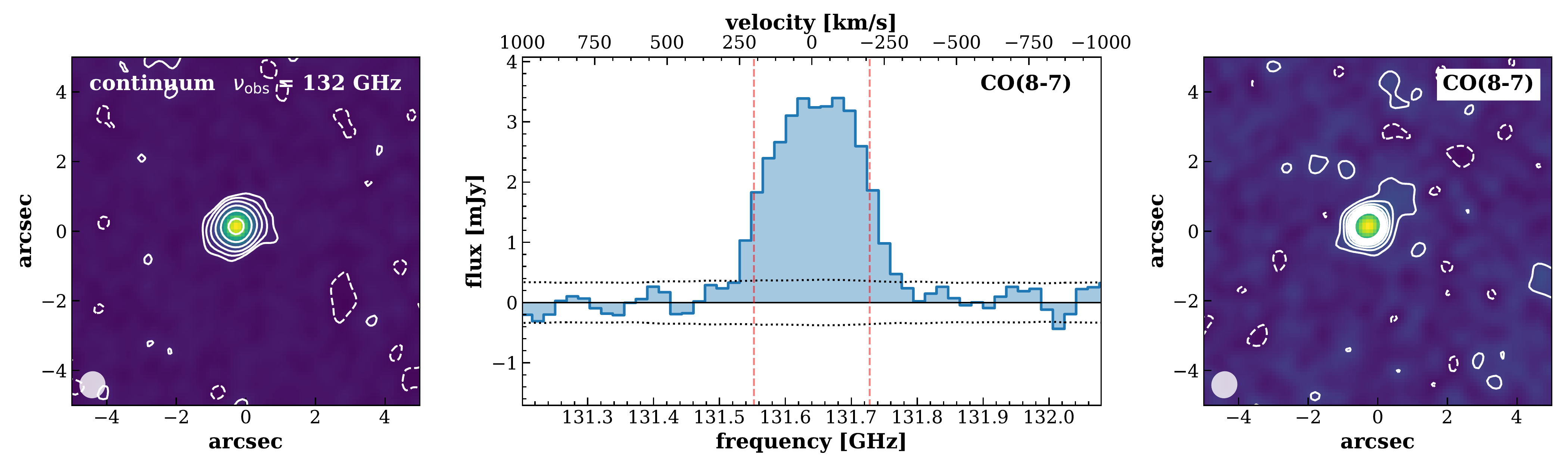}
\includegraphics[width=2\columnwidth]{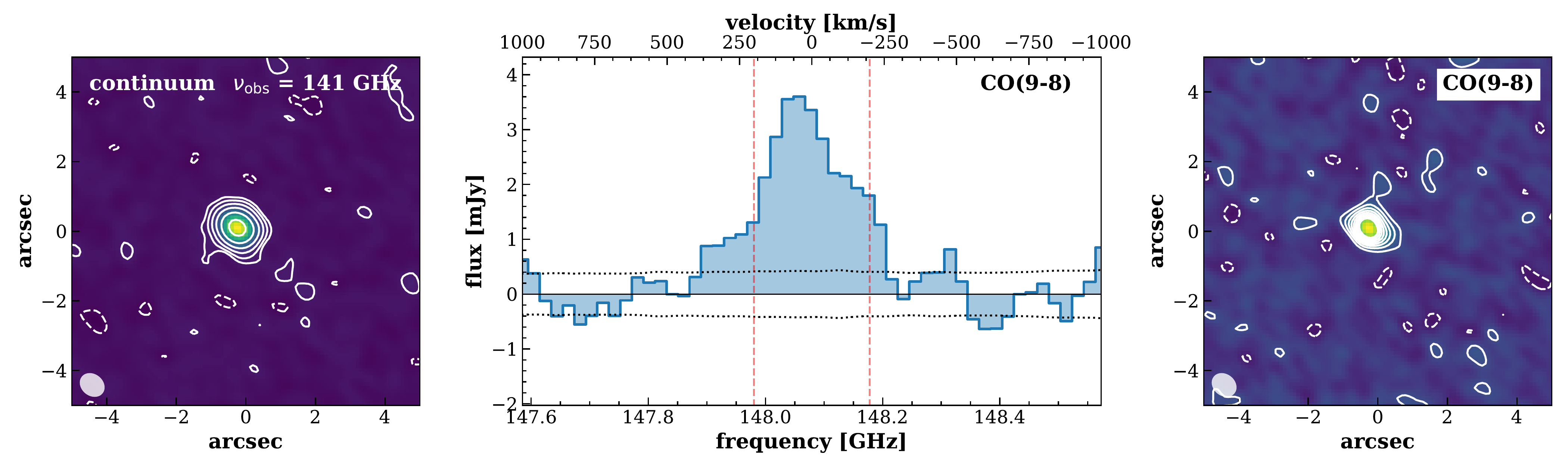}
\includegraphics[width=2\columnwidth]{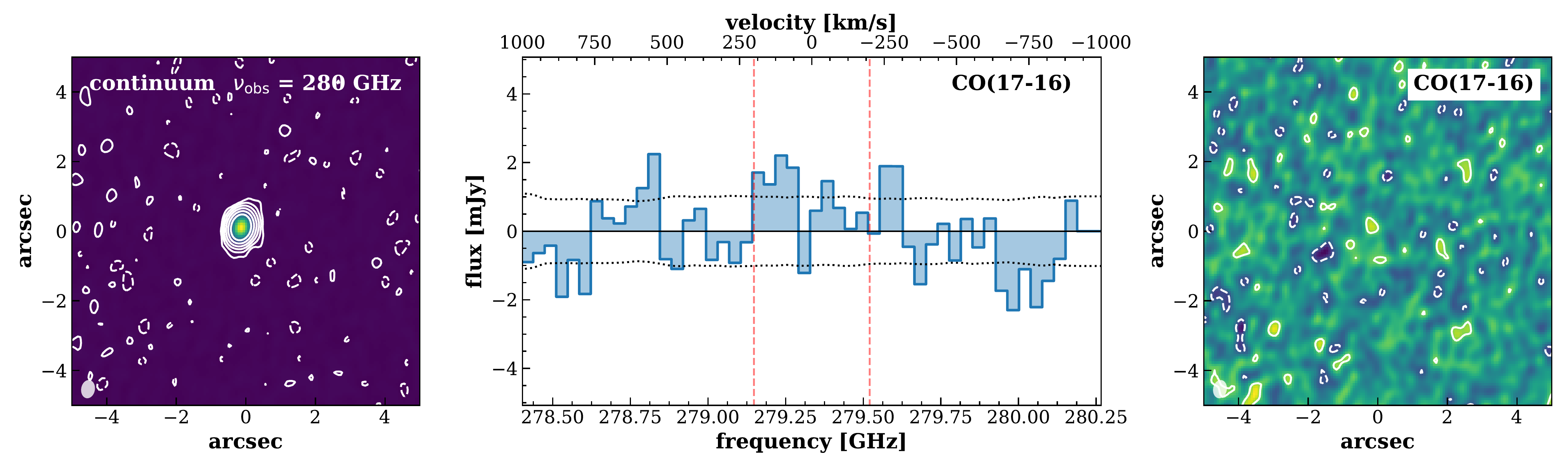}
\caption{J2310. \emph{Left panels:} continuum emission maps at 132 GHz (top),  141 GHz (middle) and  280 GHz (bottom). 
Contours are plotted at levels of $\sigma\times (-2, 2, 4, 8, 16, 32, 64)$. The rms noise level is 15 $\mu$Jy beam$^{-1}$, 15$\mu$Jy beam$^{-1}$, and 42 $\mu$Jy beam$^{-1}$, respectively for the three maps. In each map, the ALMA beam is plotted on the bottom left corner. 
\emph{Central panels}: continuum subtracted spectra extracted from a region of radius 1.5\arcsec. From top to bottom we report the spectrum of CO(8-7), CO(9-8), and CO(17-16) . All spectra are rebinned at  the same spectral resolution of 40 km s$^{-1}$ and the horizontal dotted lines indicate the noise level as a function of frequency. 
Vertical dashed red lines represent the frequency/velocity range used to extract the flux map shown in the right panels. 
\emph{Right panels:} flux maps of CO(8-7) (top), CO(9-8) (middle), and CO(17-16) (bottom). Emission is integrated in channels with $|v|<200$ km s$^{-1}$ relative to the frequency of the CO lines. Contours are shown in step of 2$\sigma$ starting from $\pm$2$\sigma$, where 1$\sigma$ is 40 mJy~beam$^{-1}$~km~s$^{-1}$, 40 mJy~beam$^{-1}$~km~s$^{-1}$, and 70 mJy~beam$^{-1}$~km~s$^{-1}$, respectively for the three molecular lines.}
\label{fig:j2310_co}
\end{figure*}

\begin{figure*}
    \centering
    {\huge J1319}\\
	\includegraphics[width=2\columnwidth]{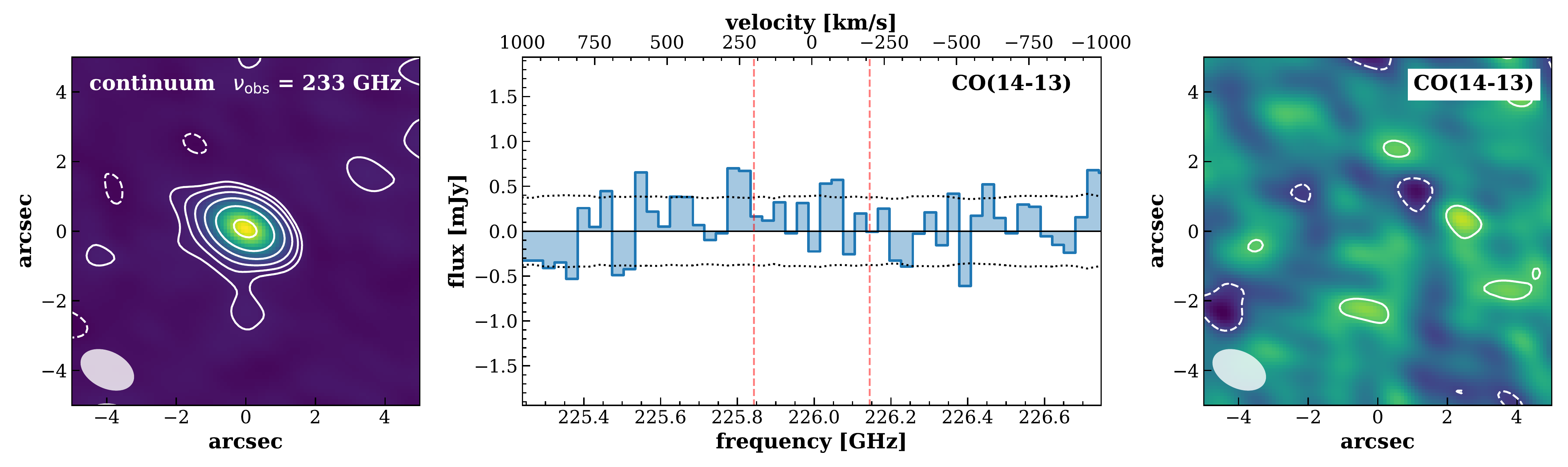}
\includegraphics[width=2\columnwidth]{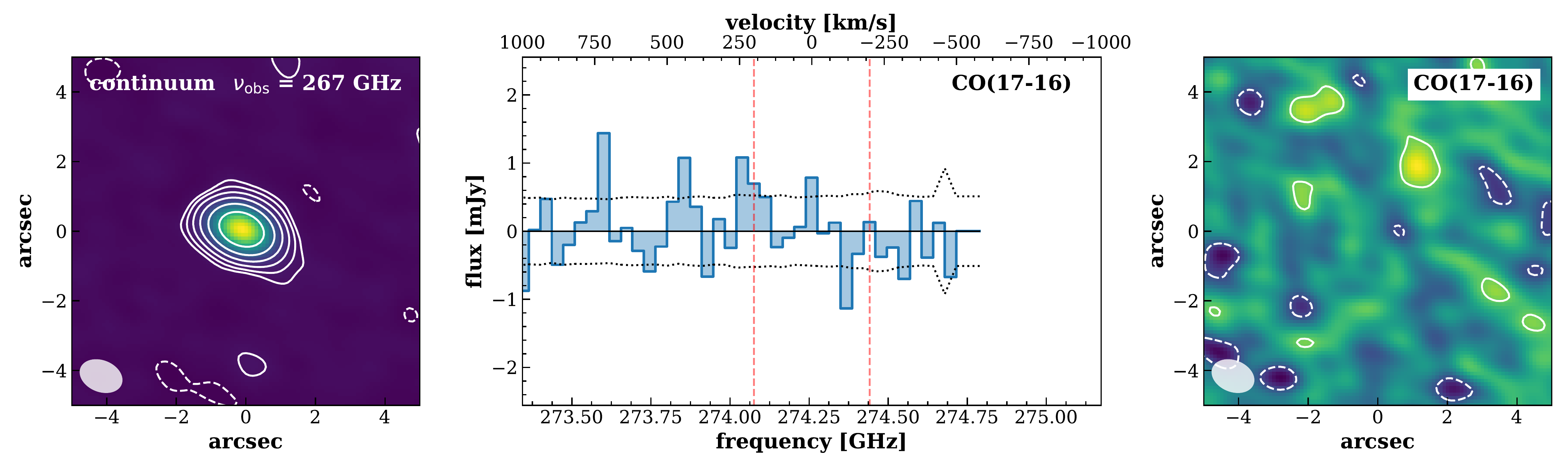}
\includegraphics[width=2\columnwidth]{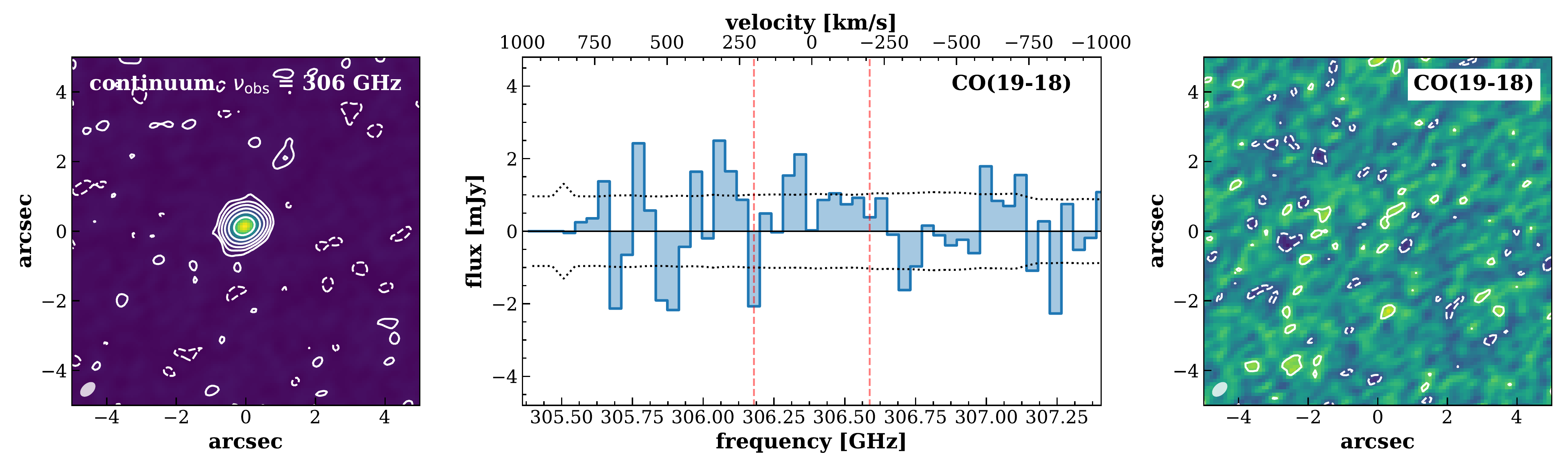}
\caption{J1319. \emph{Left panels:} continuum emission maps at 233 GHz (top),  267 GHz (middle) and  306 GHz (bottom). Contours levels are [-4, -2, 2, 4, 8, 16, 32, 64, 128] times $\sigma$. 
The 1$\sigma$ is 45 $\mu$Jy beam$^{-1}$, 43$\mu$Jy beam$^{-1}$, and 42 $\mu$Jy beam$^{-1}$, respectively for the three maps. In each map, the ALMA beam is plotted on the bottom left corner. 
\emph{Central panels}: continuum subtracted spectra extracted from a region of radius 1.5\arcsec. From top to bottom we report the spectrum of CO(14-13), J1319and J1319 CO(19-18) . All spectra are rebinned at  the same spectral resolution of 40 km s$^{-1}$ and the horizontal dotted lines indicate the noise level as a function of frequency. 
Vertical dashed red lines represent the frequency/velocity range used to extract the flux map shown in the right panels. 
\emph{Right panels:} flux maps of CO(14-13) (top), CO(17-16) (middle), and CO(19-18) (bottom). Emission is integrated in channels with $|v|<200$ km s$^{-1}$ relative to the frequency of the CO lines. Contours are shown in step of 2$\sigma$ starting from $\pm$2$\sigma$, where 1$\sigma$ is 80~mJy~beam$^{-1}$~km~s$^{-1}$, 80~mJy~beam$^{-1}$~km~s$^{-1}$, and 60~mJy~beam$^{-1}$~km~s$^{-1}$, respectively for the three molecular lines.
}
\label{fig:j1319_co}
\end{figure*}

The continuum emission is detected with high level of significance ($>10\sigma$)   in all ALMA datasets for both quasars (left panels of Figure~\ref{fig:j2310_co}~and~\ref{fig:j1319_co}). 
The ALMA coordinates of the two sources are RA=23:10:38.8994 DEC= +18.55.19.83716 (J2310) and RA=13:19:11.2879 DEC= +09.50.51.526 (J1319)  and are in agreement with those reported by  previous works \citep{Wang:2013, Shao:2017, Feruglio:2018}.  The centre coordinates of the continuum emission of both sources are consistent with those estimated from the infrared Y-band images of Hubble Space Telescope, once the astrometry of the two observations has been aligned. 


We analyse the continuum emission directly in the Fourier plane (hereafter uv-plane) by fitting the interferometric visibilities. 
We adopt a S\'ersic radial profile (with fixed index $n=1$) as a model for the continuum brightness and we assume axisymmetry to produce a 2D model image. 
We use the publicly available \texttt{GALARIO} package \citep{Tazzari:2018} to compute the visibilities of the model image by sampling its Fourier transform in the same (u,v) points sampled by ALMA. The free parameters are the total flux density $S_\nu$, the half-light radius $R_e$, the position angle (P.A., rotation on the plane of sky, defined East of North), the ellipticity (or disc inclination along the line of sight) and the ($\delta$R.A., $\delta$Dec) offsets on sky w.r.t. the observations' phase centre. Since the ($\delta$R.A., $\delta$Dec) are nuisance parameters, in the results we present here we marginalise over them. We perform the fit in a Bayesian framework, exploring the parameter space using the Markov Chain Monte Carlo (MCMC) algorithm implemented in the \texttt{emcee} package \citep{Foreman-Mackey:2013}. We assume uniform priors on the free parameters and a a Gaussian likelihood $\mathcal L \propto \exp(-\chi^2/2)$ where $\chi^2=\sum_{j=1}^{N}|V_\mathrm{mod,j}-V_\mathrm{obs,j}|^2 w_j$,
with $V_\mathrm{mod,j}$ and $V_\mathrm{obs,j}$ being the model and the observed visibilities and $w_j$ the weight associated to the $j$-th visibility point. 
We use the GPU-accelerated \texttt{GALARIO} to compute $V_\mathrm{mod}$ from the model image determined by the free parameter values.

\begin{figure}
    \centering
    
	\includegraphics[width=0.8\columnwidth]{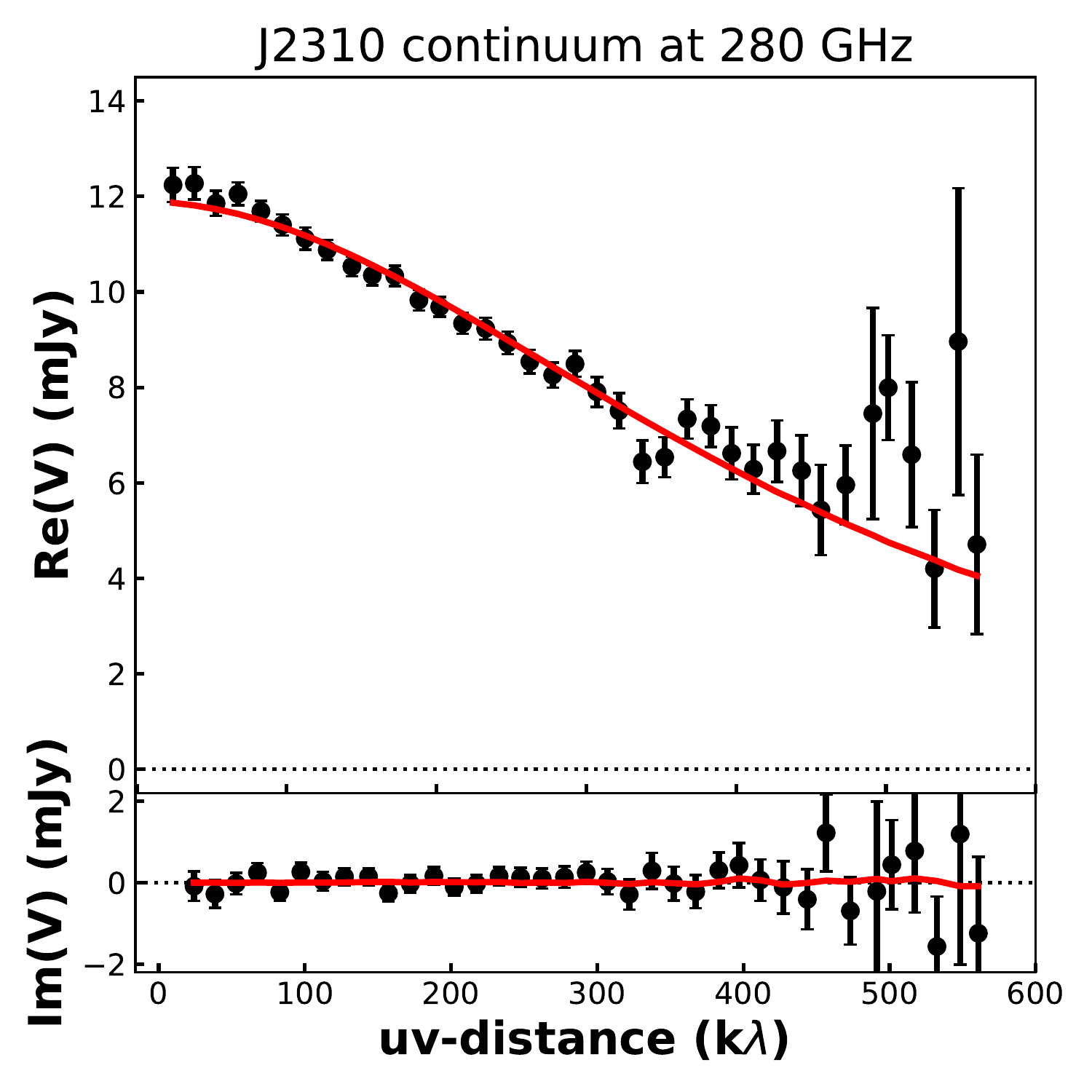}
	\caption{Comparison between the model and the observed visibilities (real and imaginary part) of the J2310 continuum emission at 280\,GHz as a function of deprojected baseline (uv-distance). For the deprojection we used the inferred inclination and P.A. The black dots are the observations, the red line shows the best-fitting model.}

\label{fig:uv}
\end{figure}

An example of the fit result for J2310(J1319) is given in Figure~\ref{fig:uv}(\ref{fig:app.uv.J1319}), where we compare the model and the observed visibilities (real and imaginary part) as a function of the deprojected baseline (uv-distance). The drop in the real part of the visibilities with the uv-distance indicates that the continuum emission is spatially resolved in the current ALMA dataset. The Figure also shows that the observed continuum profile well matches  the radial exponential model.
We report the best-fit results of both the datasets in Table~\ref{tab:cont}. More details on the fit procedure using \texttt{GALARIO} are given in Appendix~\ref{app:uv}. 
We note that the results obtained with the visibility modelling are consistent with the flux densities and de-convolved size estimate measured both  in the image plane by using the CASA task \texttt{imfit} and in the uv-plane by using the CASA task \texttt{uvmodelfit}. 

By combining all measurements from this work and literature, we infer an half light radius of $R_e=0.113\arcsec\pm0.008\arcsec~(0.66\pm0.05~{\rm kpc})$ for J2310 and of of $R_e =0.146\pm0.015\arcsec~(0.84\pm0.09~{\rm kpc})$ for J1319. This indicates that at least 50\% of the dust mass is hosted in a compact region of radius $<0.9$ kpc.

\subsection{FIR luminosities}
\label{sec:fir}
We estimate the FIR luminosities and dust masses of the two  quasars by combining our continuum data with previous ALMA, PdBI, Max-Planck-Millimeter-Bolometer (MAMBO) and Very-Large-Telescope (VLA) observations \citep{Wang:2011, Wang:2013, Feruglio:2018, Hashimoto:2018a}, whose continuum flux densities are reported in Table~\ref{tab:cont}. 

\begin{table*}
	\caption{Description of the continuum observations and derived characteristics}
	\label{tab:cont}
	\begin{tabular}{lccccccccc} 
		\hline
		 Source & $z$ & $\nu_{\rm cont}$ & $\sigma_{\rm cont}$ & S$_{\rm cont}$ & Angular & Half-light  &  Ell. & P.A & Reference \\
		 & & [GHz] & [$\mu$Jy beam$^{-1}$] & [mJy] & resolution & radius \\
		 (a) & (b) &  (c) & (d) & (e) & (f) & (g) & (h) & (i) & (j)  \\
		\hline
		\hline
		J2310 & 6.0031 & 91.5 & 5& 0.416$\pm$0.033 & 0.5\arcsec$\times$0.3\arcsec  & 0.12\arcsec$\pm$0.02\arcsec & $0.5\pm0.3$ & 138\degr$\pm$24\degr & [1]\\
		      &  & 99   & 50 & $0.40\pm0.05$ & $5.4\arcsec\times3.9\arcsec$ & -  & - & - & [3] \\
		      & & 132   & 15 & $1.39\pm0.03$ & $0.75\arcsec\times0.71\arcsec$ & $0.10\arcsec\pm0.02\arcsec$ & $<0.9$ & $100\degr\pm60\degr$ & this work\\
		      & & 141 & 15 & 1.42$\pm$0.02 & $0.81\arcsec\times0.66$ & $0.09\arcsec\pm0.02$ & $<0.2$ & $100\degr\pm60$ & this work\\
		      
		      & & 250 & 630 & 8.29$\pm$0.63 & 11\arcsec & - & - & - & [3]\\
		      & & 263 & 80 & 8.91$\pm$0.08 & 0.7\arcsec & 0.12\arcsec$\pm$0.02\arcsec & 0.2$\pm$0.1 &  162\degr$\pm$18\degr & [2] \\
		      & & 280 & 42  & 12.0$\pm$0.2 & $0.51\arcsec\times0.39\arcsec$ &  0.105\arcsec$\pm$0.016\arcsec  & $<0.12$ & $160\degr\pm40\degr$ & this work\\
		      & & 484 & 362 & 24.9$\pm$0.7 & $0.69\arcsec\times0.60\arcsec$ &  $0.15\arcsec\pm0.02\arcsec$ & $0.3\pm0.2$ & $154\degr\pm32\degr$  & [4] \\
		\hline
		J1319 & 6.1330 & 97 & 80 & $0.31\pm0.08$ & $\sim$3.5\arcsec & - & - & - & [2] \\
		      & & 233   & 45 & $3.89\pm0.12$ & $1.59\arcsec\times1.03\arcsec$ & $0.13\arcsec\pm0.02\arcsec$ & $<0.2$ & 110\degr$\pm$70\degr & this work \\
		      &  & 250 & 650 & 4.20$\pm$0.65 &  $\sim11\arcsec$ & - & - & - &  [2]\\
		      & & 258 & 100 & 5.23$\pm$0.10 & $\sim0.7\arcsec$ & $0.19\arcsec\pm0.01\arcsec$ & $0.12\pm0.08$ & $121\degr\pm148\degr$ & [3] \\
		      & & 267 & 43 & 6.03$\pm$0.07 & $1.40\arcsec\times0.97\arcsec$ & $0.13\arcsec\pm0.02\arcsec$ & $<0.2$ &  $70\degr\pm70\degr$ & this work \\
		      &  & 306 & 42 & 7.39$\pm$0.11 &$0.51\arcsec\times0.38\arcsec$ & $0.139\arcsec\pm0.005\arcsec$ &   $0.12\pm0.07$ & $55\degr\pm11\degr$ & this work \\
		      & & 347 & 33 & 9.68$\pm$0.12 & $0.78\arcsec\times0.51\arcsec$ & $0.143\arcsec\pm0.008\arcsec$ & $0.17\pm0.05$ & $55\degr\pm20\degr$ & this work \\

		 \hline


	\multicolumn{10}{l}{%
  \begin{minipage}{0.95\linewidth}%
    \footnotesize 
    NOTE -- Col.(a): object name. 
		Col.(b): redshift. 
		Col.(c): observed frequency. 
		Col.(d): rms on the continuum. 
		Col.(e): continuum flux density. 
		Col.(f): angular resolution.
		Col.(g): half-light radius $R_{e}$ of  continuum emission;
		We note that in literature the galaxy size is usually reported in term of either FWHM, if a gaussian profile has been used as model, or radius scale $r_{d}$, in the case of an exponential profile model. $R_{e}$ is related to FWHM and $r_{d}$ as $R_e \approx \sqrt{-2\ln(0.5)}FHWM/2.355$ and $R_e \approx 1.67835r_{d}$, respectively.
		Col.(h): ellipticity.
		Col.(i): position angle.
   		Col.(j) reference; [1] \cite{Feruglio:2018} [2] \cite{Wang:2013} [3] \cite{Wang:2011} [4] \cite{Hashimoto:2018a}. 
		
  \end{minipage}%
}\\
		\end{tabular}\\

\end{table*}

The spectral-energy-distribution (SED) of dust emission can be represented with a  modified black-body function, which is also dubbed grey-body law, given by:
\begin{equation}
\label{ref:dust_classic}
S^{\rm obs}_{\nu_{\rm obs}} = S^{\rm obs}_{\nu/(1+z)} =
\frac{\Omega}{(1+z)^3}B_{\nu}(T_{\rm dust})(1-e^{-\tau_\nu}),
\end{equation}
where $S^{\rm obs}_{\nu_{\rm obs}}$ is the flux density measured at the observed frequency $\nu_{\rm obs}={\nu/(1+z)}$, $\Omega$ is the solid angle subtended by the galaxy\footnote{ The solid angle $\Omega$ can be also written as $\Omega=(1+z)^4 A_{\rm galaxy}D_{\rm L}^{-2}$ where $A_{\rm galaxy}$ and $D_L$ are the  surface area and luminosity distance of the galaxy, respectively.}, $B_{\nu}(T_{\rm dust})$ is the black-body emission at the dust temperature $T_{\rm dust}$, and $\tau_\nu$ is the dust optical depth.
The optical depth can be expressed in the form
\begin{equation}
\tau_\nu = \Sigma_{\rm dust} k_{\nu} = \Sigma_{\rm dust} k_0 \left(\frac{\nu}{{\rm \nu}_0}\right)^{\beta},
\end{equation}
where $\Sigma_{\rm dust}$ and $k_{\nu}$ are the surface mass density of dust and  dust opacity \citep{Draine:1984}, respectively. The $\Sigma_{\rm dust}$ can be expressed in term of total dust mass ($M_{\rm dust}$) and the physical area of the galaxy ($A_{\rm galaxy}$): $\Sigma_{\rm dust}$ = $M_{\rm dust}/A_{\rm galaxy}$ \citep{da-Cunha:2013}; the dust opacity $k_{\nu}$ depends on the emissivity index $\beta$, mass absorption coefficient k$_0$ and $\nu_0$. The latter two terms, k$_0$ and $\nu_0$, are usually assumed from either observations (e.g. \citealt{Alton:2004, Beelen:2006})  or dust models (e.g. \citealt{Bianchi:2007}). In this work we assume k$_{\nu}$ = 0.45$\times(\nu$/250 GHz)$^{\beta}$ cm$^{2}$ g$^{-1}$ \citep{Beelen:2006}.

In equation~\ref{ref:dust_classic} we must also consider the contribution of the CMB emission since, at $z=6$, the CMB temperature may be comparable (or even higher) to that of dust, thus affecting  observations and measurements.
An extensive discussion about the CMB effect on the dust emission has been already presented by \cite{da-Cunha:2013}; in this section we only provide a quick review of their main results. 

In addition to the galaxy radiation, CMB photons contribute to increase the dust temperature and boost the observed dust emission. The effect on the dust temperature is:
\begin{equation}
\label{eq:T_dust}
T_{\rm dust}(z) = ((T_{\rm dust})^{4+\beta} + {T_{0}}^{4+\beta}[(1+z)^{4+\beta}-1])^{\frac{1}{4+\beta}}
\end{equation}
where $T_{0}$ is the CMB temperature at $z=0$, i.e. $T_{0}$= 2.73 K \citep{Fixsen:1996}. 
On the other hand, the CMB emission at high-$z$ is a strong background against which the dust continuum emission is observed, thus reducing its detectability. The observed SED shape of dust emission can thus be expressed as:
\begin{equation}
\label{eq:dust}
S^{\rm obs}_{\nu/(1+z)}  =  \frac{\Omega}{(1+z)^{3}}\left[B_{\nu}(T_{\rm dust}(z))-B_{\nu}(T_{\rm CMB}(z))\right](1-e^{-\tau_\nu})
\end{equation}
where $T_{\rm dust}(z)$ is given by equation~\ref{eq:T_dust} and $B_{\nu}$(T$_{\rm CMB}$(z)) is the black-body emission of CMB at the temperature $T_{\rm CMB}$(z)= T$_{0}(1+z)$.

We use equation~\ref{eq:dust}  to fit the flux continuum densities measured in J2310 and J1319. We estimate the solid angle subtended by each galaxy from the continuum emission size inferred in Section~\ref{sec:continuum.emission}.
and then explore the three free parameters, $M_{\rm dust}$, $T_{\rm dust}$, and $\beta$, by using a MCMC algorithm  to estimate the posterior probability distribution for the 3-dimensional parameter space that defines our SED model. We employ uniform distributions for the priors, but we force $30~{\rm K}<T_{\rm dust}<90~{\rm K}$ and $1.0<\beta<2.0$ since these are the typical ranges observed in star-forming galaxies and quasars \citep{Beelen:2006}. 

\begin{figure}
	\includegraphics[width=\columnwidth]{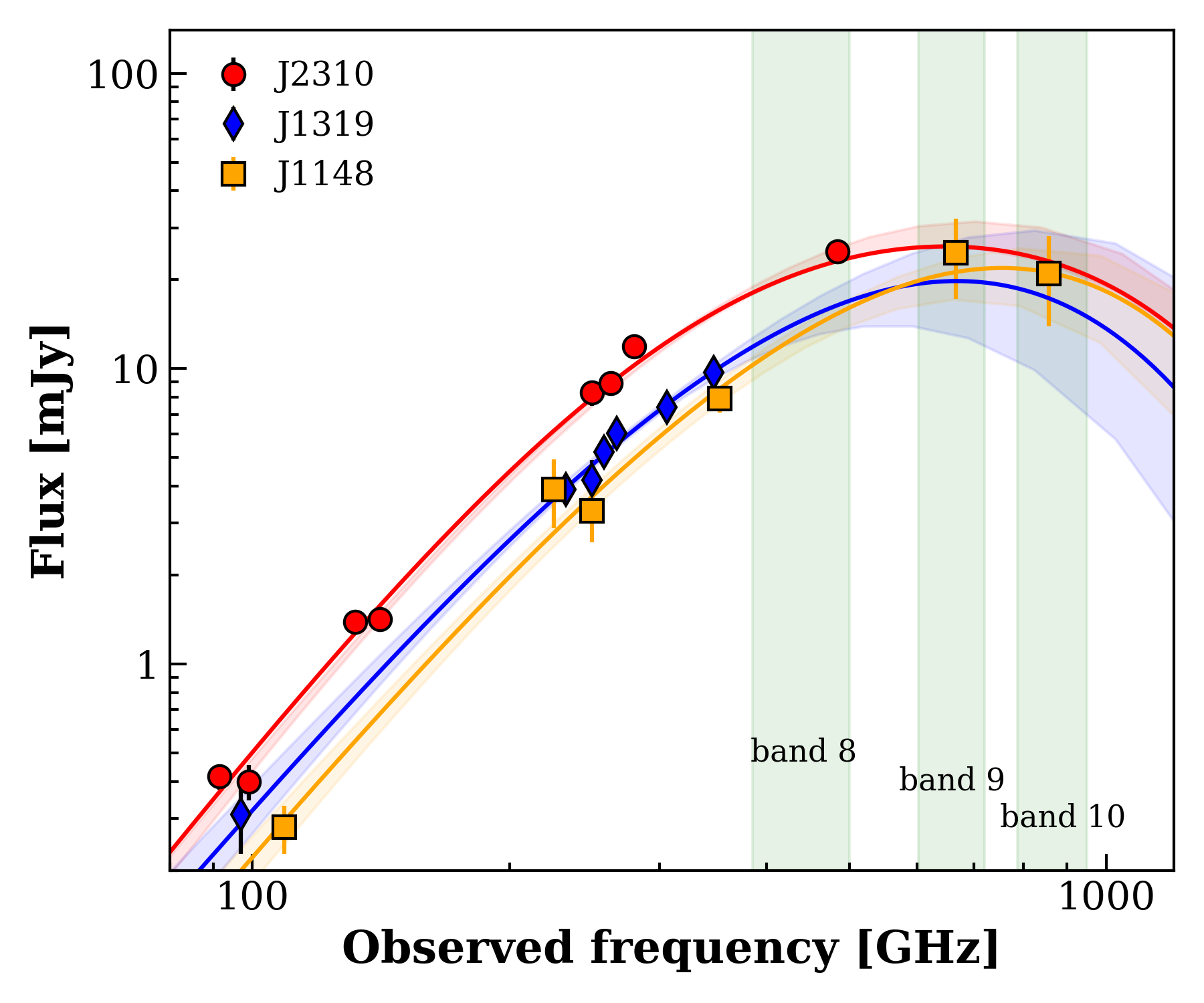}
    \caption{Spectral energy distribution of J2310 (red), J1319 (blue), and J1148 (orange). The observations are reported as circle, diamonds, and square marks, while the lines indicate the SED  best-fitting models. The vertical shaded green region shows the range of frequency covered by ALMA band 8, 9, and 10, which are fundamental to constrain dust properties in high-$z$ sources (see text). }
    \label{fig:sed}
\end{figure}

\begin{figure}
	\includegraphics[width=0.81\columnwidth]{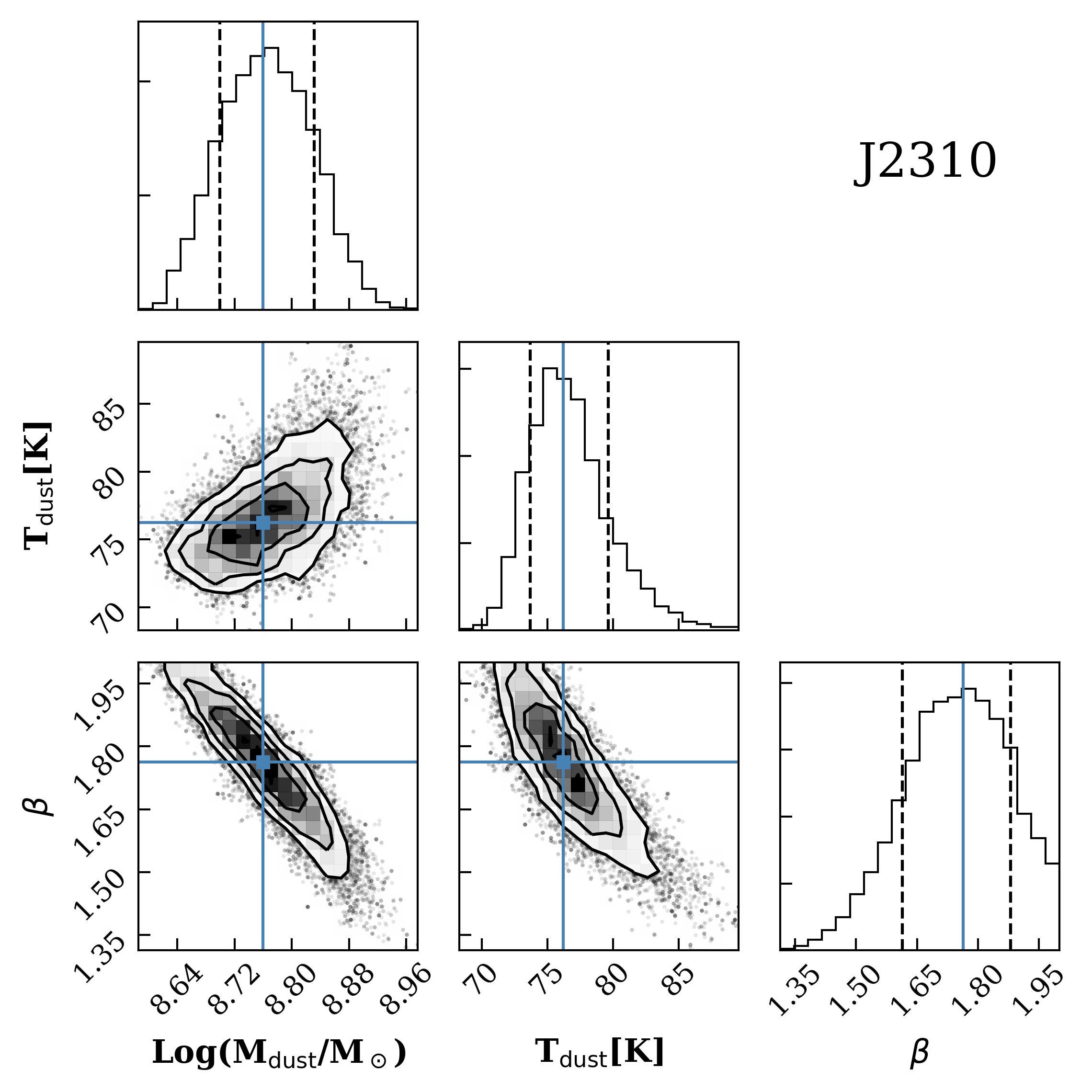}
	\includegraphics[width=0.81\columnwidth]{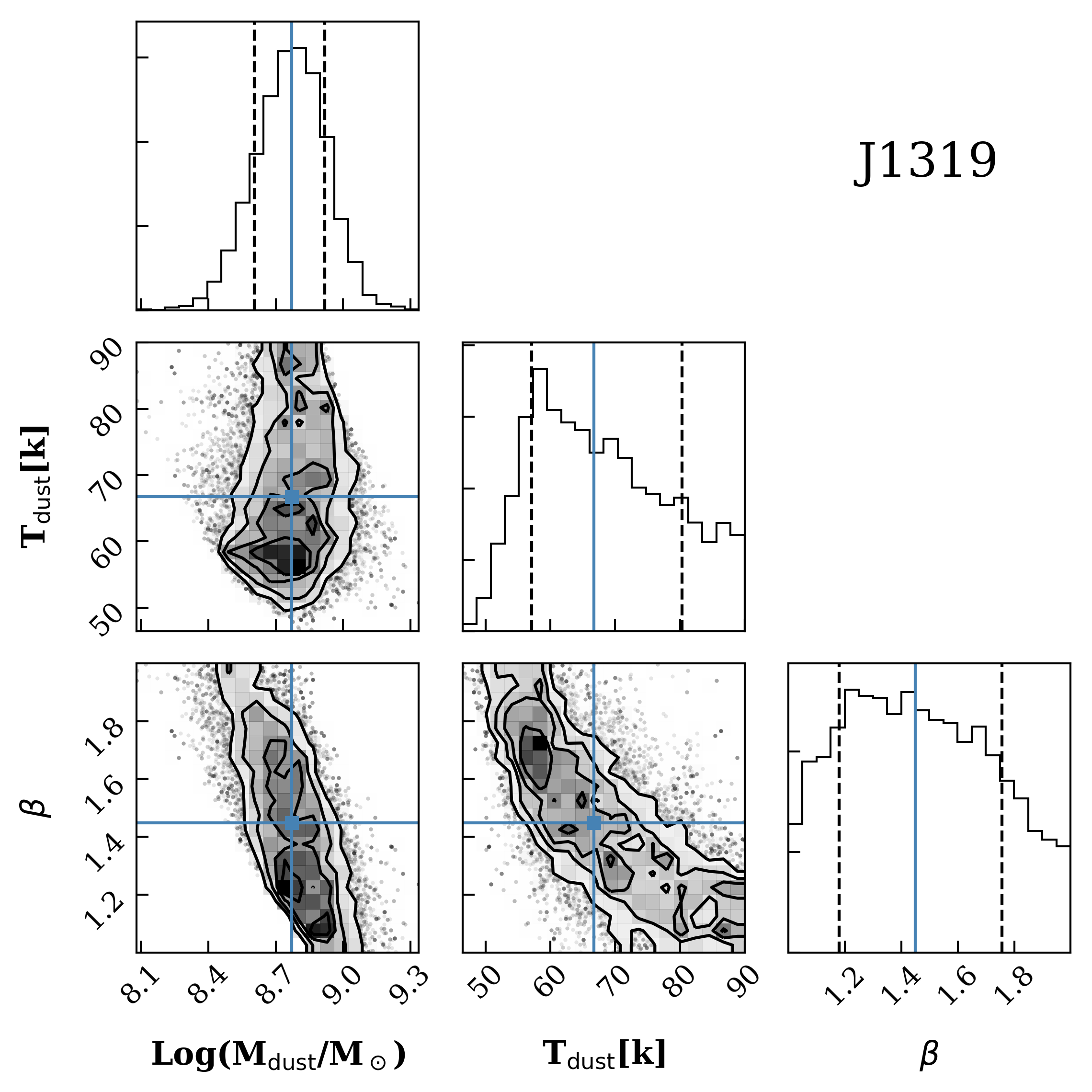}
	\includegraphics[width=0.81\columnwidth]{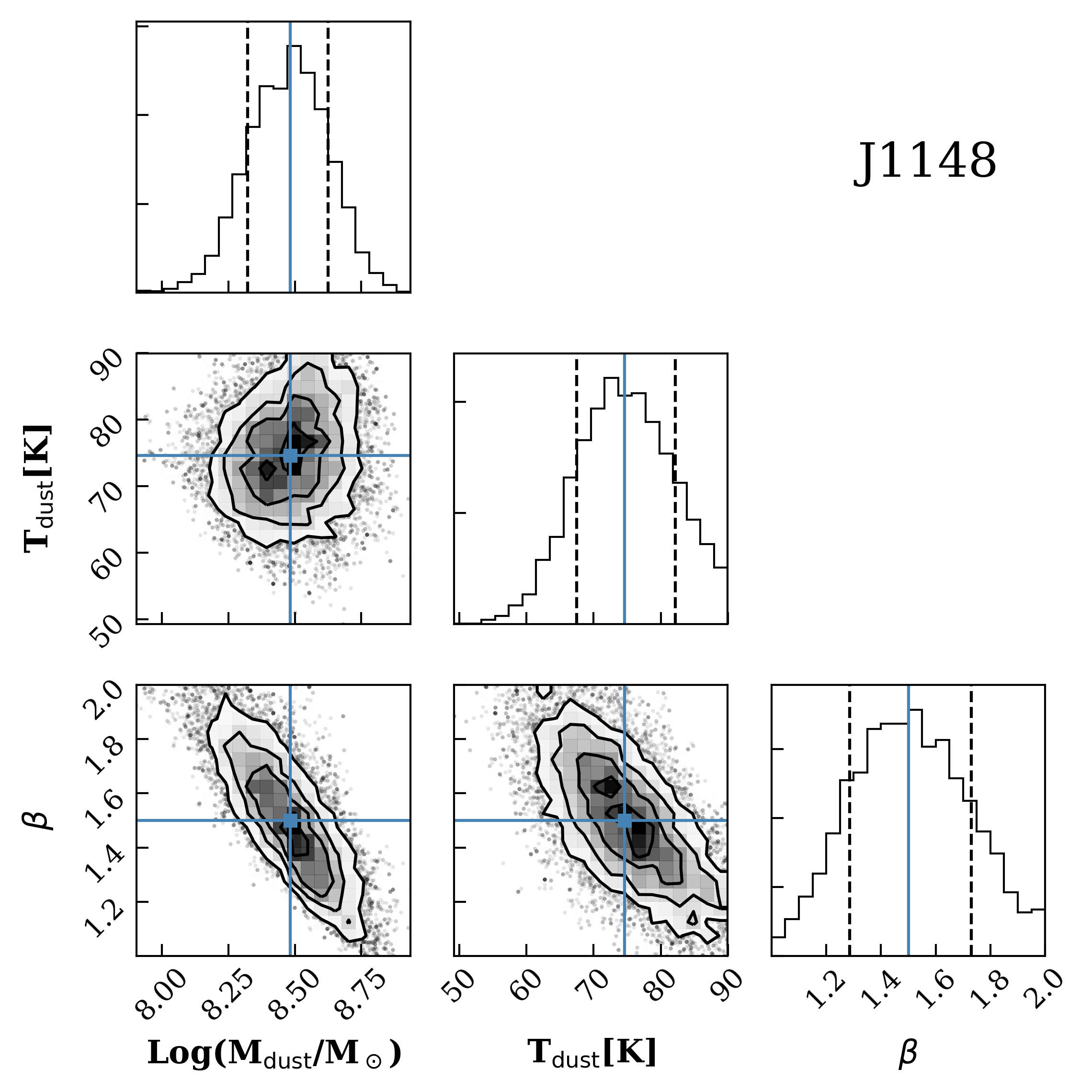}

    \caption{Corner plots showing the three dimensional posterior probability distributions of $M_{\rm dust}$, $T_{\rm dust}$, and $\beta$ for J2310 (top panel), J1319 (middle panel), and J1148 (bottom panel). Blue lines indicate the best-fit parameter, while the dashed lines show the 16\% and 84\% percentiles for each parameter. The lack of high-frequency observations does not allow us to constrain the $T_{\rm dust}$ and $\beta$ of J1319.}
    \label{fig:corner}
\end{figure}

Figure~\ref{fig:sed} shows the results of the SED model fitting, 
Table~\ref{tab:prop} reports the best-fitting results, and Figure~\ref{fig:corner} shows the confidence contours for the three free parameters obtained from a MCMC with 50 chains and 3000 trials. 
While the dust masses are well pinned down in both sources with errors smaller than 20\%, the emissivity index and dust temperature are constrained only in J2310, where the ALMA \mbox{band-8} observations are present.  
We note that our best-fitting $T_{\rm dust}$ values for J2310 and J1319 are higher than the dust temperatures ($T_{\rm dust}=40-50$~K) inferred in previous works \citep{Wang:2013,Shao:2019}. 
This discrepancy can be explained by the different assumptions adopted on the dust opacity when modelling the SED. Indeed, previous works have assumed an optically thin ($\tau_\nu<<1$) modified blackbody profile to fit the continuum observations. Here, we are instead also considering  dust emission attenuation. In fact, the dust masses and continuum sizes of our AGN host galaxies provide $\tau_\nu>0.5$ at $\nu_{\rm obs}>300$~GHz, thus making the optically thin approximation not valid at higher frequencies. 
For comparison in Appendix~\ref{app:thin} we report the best-fitting results in the optical thin assumptions, which are in agreement with those reported by previous studies \citep{Wang:2013,Shao:2019}.

The shadowed red and blue regions in Figure~\ref{fig:sed} give an indication of the $\beta$ and $T_{\rm dust}$ degeneration associated to the best-fitting models. While the shaded regions shrink at lower frequencies ($<400$ GHz), the uncertainties enlarge close to the peak of the curve. This indicates that high frequency observations, as that in band 8 for J2319, are fundamental to constrain the properties of the dust in the distant Universe. Observations in ALMA band 8, 9 and 10, which are the highest frequency bands in the baseline ALMA project, are thus crucial to compute the dust temperature in the first billion years of the Universe. 

We perform the SED fitting of J1148 as well by using the same assumptions made for the other two quasars and taking into account the CMB contribution.
We retrieve from the literature  all  flux continuum densities \citep{Bertoldi:2003, Walter:2003,Riechers:2009, Cicone:2015} at the wavelength $\lambda_{\rm rest}>50~\mu$m where the emission is powered mainly by star-formation activity in the host galaxy and the contribution from the AGN is negligible \citep{Leipski:2013}. 
Given the presence of a serendipity source at 10.5\arcsec, we give less weight to those continuum measurements of J1148 that may be contaminated by the emission of a serendipity source \citep{Cicone:2015}.
The results of the Bayesian fit is reported in Figure~\ref{fig:sed}, Figure~\ref{fig:corner}, and Table~\ref{tab:prop}. 

We estimate the FIR luminosities by integrating the best-fit models from 8~$\mu$m to 1000~$\mu$m rest-frame for the three quasars. 
Despite the comparable FIR luminosities ($L_{\rm FIR}\sim10^{13}$~\lsun), from the SED fitting it results that the dust temperature of J1148 is 2$\sigma$ higher than that inferred for J2310, possibly due to a stronger radiation field in the former.
For all quasars we turn the FIR luminosities into SFR by using the relations from \cite{Kennicutt:2012}.  All quasars have SFR $> 1000$ \sfr (see Table~\ref{tab:tab3}).

\begin{table}
	\centering
	\caption{Results of the SED fitting (assuming $k_0=0.45~\rm cm^{2}~g^{-1}$ and $\nu_0=250~\rm GHz$)
	and CO observations.}
	\label{tab:prop}
	\begin{tabular}{lccc } 
		\hline
				 & J2310 & J1319 & J1148 \\
		\hline
		\hline
		\multicolumn{4}{c}{Dust emission}\\
		\hline
		\hline
		Log(M$_{\rm dust}$/M$_\odot$) & $8.75\pm0.07$ & $8.8\pm0.2$  & $8.5\pm0.1$ \\
		T$_{\rm dust}$ [K] & $76\pm3$  & $66^{+15}_{-10}$  & $75\pm8$\\
		$\beta$ & $1.8\pm0.1$ & $1.5\pm0.3$ & $1.5\pm0.2$\\
		L$_{\rm FIR}$ [10$^{13}$ \lsun] & $2.5\pm0.4$ & $1.6^{+1.3}_{-0.6}$  & $2.1\pm0.7$\\

	\hline
		\hline
		\multicolumn{4}{c}{CO emission [$10^{8}$ \lsun]}\\
		\hline
		\hline
	    CO(1-0)& - & - & $<$0.72$^d$ \\
	    CO(2-1)& $0.213\pm0.002^{a}$ & - &  $0.125\pm0.010^e$\\
	    CO(3-2) & - & - & $0.39\pm0.04^f$ \\
	    CO(6-5) & $5.5\pm0.5^{b}$ & $1.6\pm0.3^c$ &  $2.5\pm0.3^d$ \\
	    CO(7-6) & - & - & $2.9\pm0.3^g$ \\
	    CO(8-7) & $7.1\pm0.6$ & - & - \\
	    CO(9-8) & $7.5\pm0.9$ & - & - \\
	    CO(14-13) & - & $<1.8^\ddag$  & - \\
	    CO(17-16) & $<8.1^\ddag$  & $<2.6^\ddag$ & $4.9\pm1.1$$^{h\dag}$\\
	    CO(19-18) & -  & $<3.2^\ddag$  & - \\
	    	\hline
		\hline
		
		\multicolumn{4}{l}{%
  \begin{minipage}{0.95\linewidth}%
    \footnotesize 
    REFERENCE: $^{a}$\cite{Shao:2019}; $^{b}$\cite{Feruglio:2018}; $^c$\cite{Wang:2013}; $^d$\cite{Bertoldi:2003a}; $^e$\cite{Stefan:2015}; $^f$\cite{Walter:2003}; $^g$\cite{Riechers:2009}; $^h$\cite{Gallerani:2014}.
    NOTE: $^\ddag$Upper limits are estimated in a aperture of $1.4\arcsec\times1.4\arcsec$, and spectral width of $\Delta v = 400$ \kms\ for J2310 and $\Delta v = 500$ \kms\ for J1319; $^\dag$ The CO(17-16) luminosity of J1148 accounts for the possible contamination by OH$^+$. See \citealt{Gallerani:2014} and Sec. \ref{J1148comparison} for further details. 
		
  \end{minipage}}	
	\end{tabular}

\end{table}

\subsection{High-J CO emission}
\begin{figure*}
 	\includegraphics[width=8.8cm]{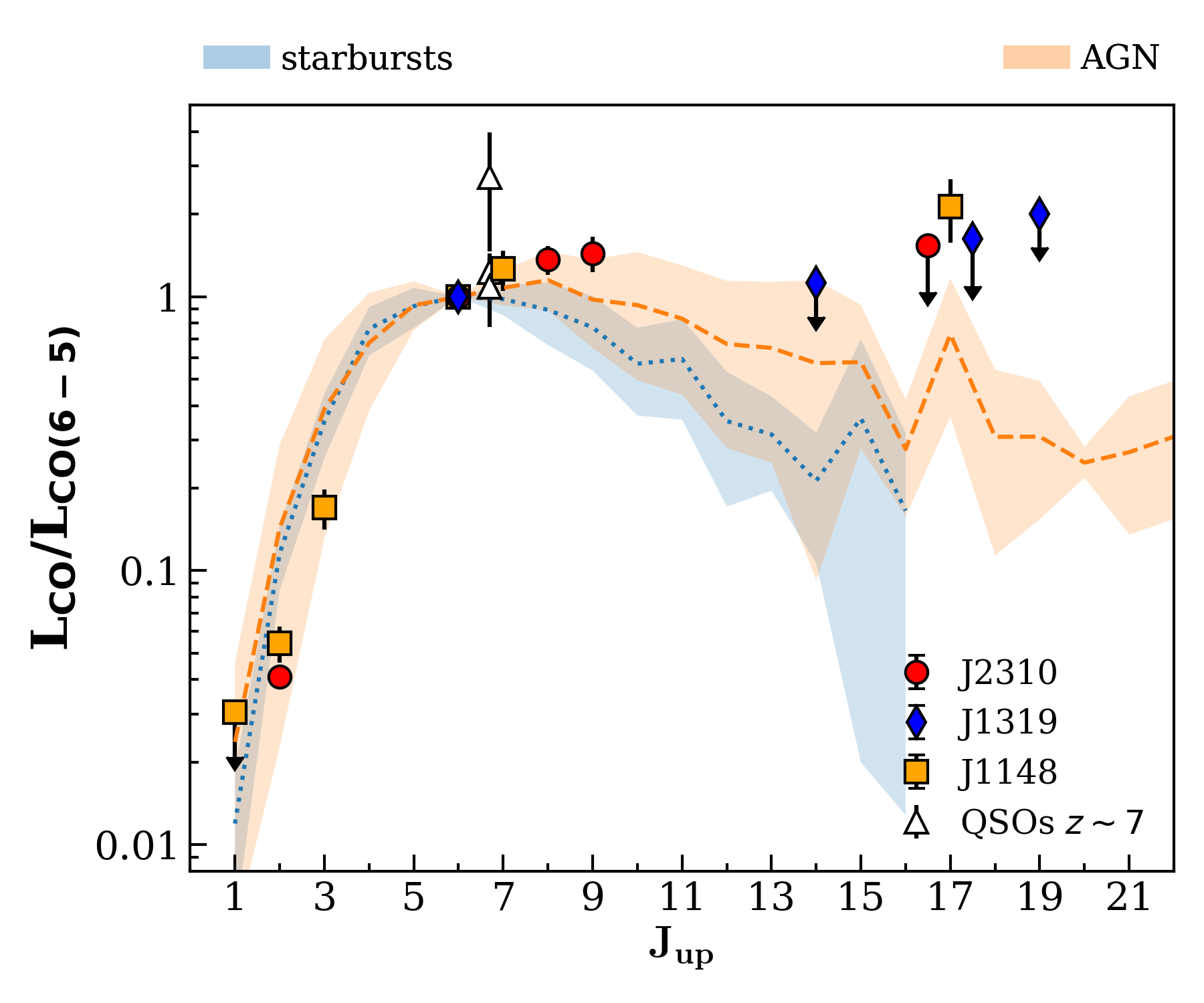}
 	\includegraphics[width=8.8cm]{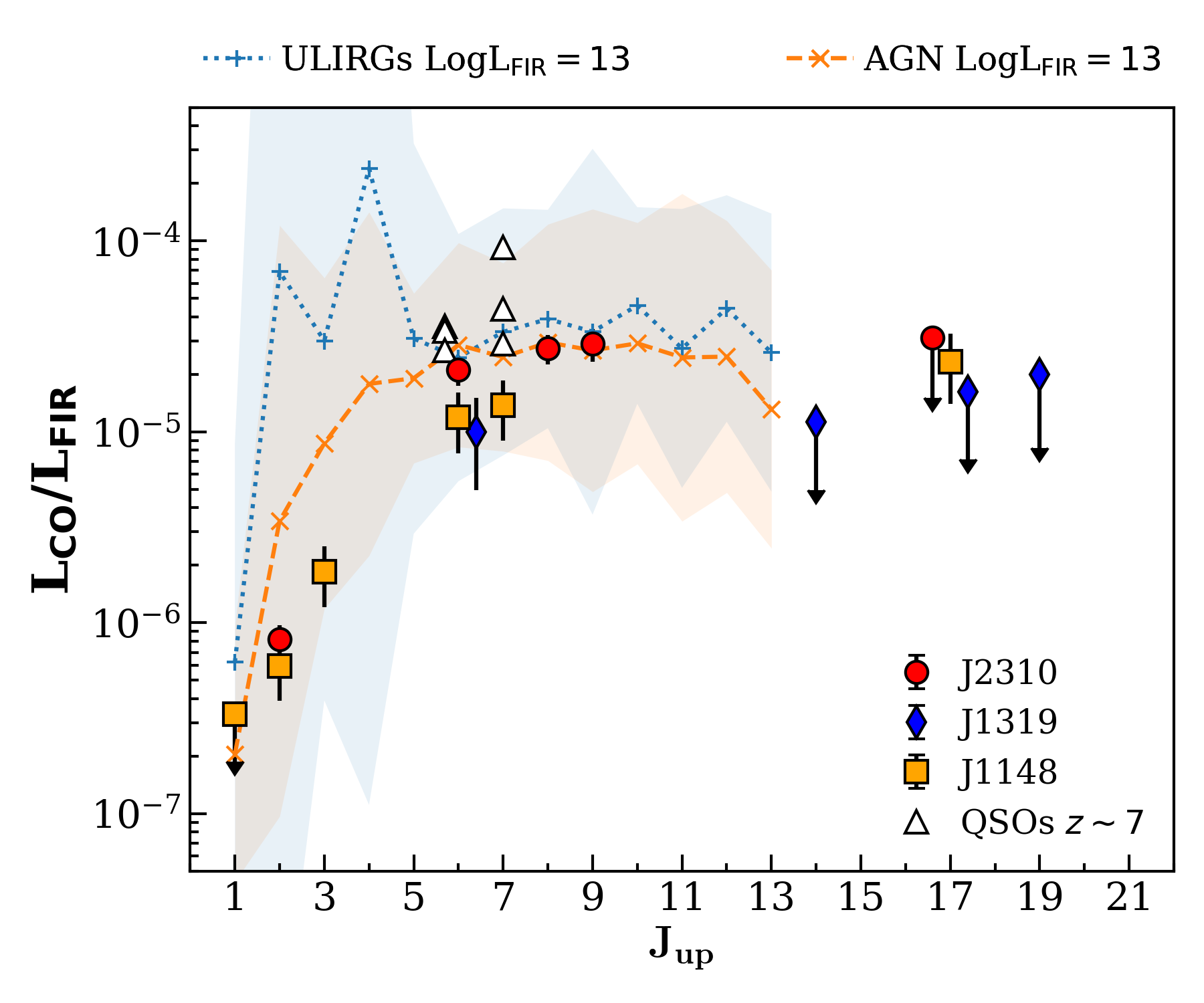}
    \caption{\textbf{Left panel}:  Average  CO SLED of local AGN  (dashed red line) and starburst galaxies (dotted blue line) from literature \citep{Rosenberg:2015,Mashian:2015,Mingozzi:2018}. The list of the local galaxies is reported in Table~\ref{tab:local_co_sled}. CO SLEDs are normalised to the CO(6-5) line. The shaded regions show the variance in each observed line. Orange, red, and blue marks indicate the CO measurements of J1148, J2310, and J1319, respectively. \textbf{Right panel}: CO SLED normalised by FIR luminosity of individual quasar: J1148 (orange), J2310 (red), and J1319 (blue). The dashed and dotted lines indicates the average L$_{\rm CO}$/L$_{\rm FIR}$ ratio estimated from the each \jup\ relations Log(L${^\prime}_{\rm CO}$)=$\alpha {\rm Log}({\rm L_{\rm FIR}})+\beta$ inferred for  AGN and ULIRGs, respectively \citep{Kamenetzky:2016}}
     \label{fig:cosled}
\end{figure*}

\subsubsection{CO measurements in J2310}

While the continuum emission is detected in all ALMA datasets of J2310, the detection of \jup~$>7$ CO lines is limited only to the two lower transitions, CO(8-7) and CO(9-8) (Figure~\ref{fig:j2310_co}; see also Li et al. in prep.).
Both CO lines have a line width of 380 km~$\rm s^{-1}$  and a beam-deconvolved size of $0.50\arcsec\pm0.12\arcsec$ ($\sim3$ kpc at $z=6$). We measure  integrated flux densities of $1.49\pm0.13$ Jy km s$^{-1}$ for CO(8-7) and $1.41\pm0.16$ Jy km s$^{-1}$ for CO(9-8) that correspond to a luminosity of $(7.1\pm0.6)\times10^{8}$ \lsun\ and $(7.5\pm0.9)\times10^{8}$ \lsun, respectively  (Table~\ref{tab:prop}).  

The flux map of the CO(17-16) emission, which has been obtained by integrating the datacube within $\pm$200 km~$\rm s^{-1}$ from the expected frequency of the \jup~$=17$ CO line, shows a marginal detection of $2 \sigma$ at the location of the quasar. To be conservative we estimate a $3\sigma$ upper limit on its flux density in an aperture of $1.4\arcsec\times1.4\arcsec$\footnote{Current ALMA observations of J2310 and J1319 have different angular resolutions. To compare the CO upper limits, the flux densities have been estimated from the same size aperture as large as the ALMA beam of the poorest angular resolution dataset ($\sim1.4\arcsec$).}, yielding $L_{\rm CO(17-16)}<8.1\times10^{8}$ \lsun. 

\subsubsection{CO measurements in J1319}

At the location of the quasar J1319 all three ALMA observations do not reveal any signature of \jup~$>7$ CO emission. The right panels of Figure~\ref{fig:j1319_co} show the flux maps obtained by collapsing the datacubes over a channel width of 500 km s$^{-1}$ that is comparable to the line widths observed in the CO(6-5) and [\ion{C}{ii}] lines \citep{Wang:2013, Shao:2017}. In Table~\ref{tab:prop} we report the $3\sigma$ upper limit,  measured in an aperture size as large as $1.4\arcsec\times1.4\arcsec$,  on each CO transition analysed in this work.

\section{CO SLED}\label{sec:sled}

In this section, we compute the CO SLED resulting from the ALMA data presented in this work and from the other CO measurements in  literature. We  then compare our findings with the CO SLEDs observed in low-$z$ starburst galaxies and AGN, and the CO SLEDs measured in other $z\sim 6$ quasars.

\subsection{CO(6-5)-normalised CO SLED}

In the left panel of Fig.~\ref{fig:cosled}, we show the CO SLED (normalised to the CO(6-5) line) of J2310, J1319, J1148 \citep{Bertoldi:2003a, Riechers:2009, Gallerani:2014, Stefan:2015}, and the three $z\sim 6$ quasars by \cite{Venemans:2017a}. 
The blue and red shaded regions represent the averaged CO(6-5)-normalised CO SLEDs from low-$z$ starburst galaxies and AGN, respectively \citep[Table~\ref{tab:local_co_sled},][]{Rosenberg:2015,Mashian:2015, Mingozzi:2018}. 
Despite the large uncertainties, the CO SLED of the AGN population is characterised by a slightly different shape with respect to the one of starburst galaxies. At high-J (\jup~$\geq 13$) the starburst CO SLED is less excited and the CO emission steadily declines at \jup~$>8$, while the AGN CO SLED seems to reach the peak at $8\le$~\jup~$\le10 $ and declines afterwards.

The CO luminosities and upper limits measured for J2310 and J1319 are consistent with both AGN and starburst CO SLEDs. Although the L$_{\rm CO(8-7)}$/L$_{\rm CO(6-5)}$ and L$_{\rm CO(9-8)}$/ L$_{\rm CO(6-5)}$  ratios measured in J2310 are slightly higher relative to the typical ratios observed in starburst galaxies, they are still consistent (within $1\sigma$) with the CO SLED shape of both populations (AGN and starburst galaxies). The same is true for the three $z\sim 6$ quasars by \cite{Venemans:2017a} that have $1~\lesssim~L_{\rm CO(7-6)}/L_{\rm CO(6-5)}~\lesssim~4$, consistently with low-$z$ observations. 

For what concerns J1148, at lower \jup\ ($<7$), the CO(6-5)-normalised SLED is also similar to that observed in local AGN and starburst populations. However this quasar is characterised by an exceptionally strong CO(17-16) emission line that is $>3\sigma$ larger than the averaged CO SLED of both populations.  The CO ratio is also $\sim1\sigma$ higher than the upper limits on  the CO(17-16)/CO(6-5) ratios estimated for J2310 and J1319.

We further discuss the origin of the discrepancy between J1148 and other $z\sim 6$ quasars in Sec. \ref{discussion}. Here, we mention that the CO(17-16) emission detected in J1148 may be contaminated by the presence of several flanking OH$^{+}$ lines falling within few hundreds \kms\ of the CO line. These lines have been  observed in local nearby AGN host galaxies \citep{Hailey-Dunsheath:2012,Gonzalez-Alfonso:2013} and  seem to be more luminous in presence of fast outflowing gas and XDRs  \citep{Gonzalez-Alfonso:2013,Gonzalez-Alfonso:2018}. By fitting with a double Gaussian profile the emission detected in J1148, \cite{Gallerani:2014} estimate that $\sim40\%$ of the measured luminosity can be contaminated by OH$^{+}$ line emission.
We thus perform, both in J2310 and J1319,  a blind line search \citep[see][for details]{Carniani:2017,Carniani:2017a} to look for possible OH$^{+}$+CO emissions at the location and redshift of the two quasars, but we do not find any emission with an intensity $>2\sigma$.

To interpret the results found in J2310 we have repeated the analysis done in \citealt{Gallerani:2014} by adopting the PDR/XDR models by \citealt{Meijerink:2005} and \citealt{Meijerink:2007}.
The same analysis is hampered in J1319 given that, in this source, only the CO(6-5) line has been detected.  We found that a PDR model alone can fairly explain the intensity of the CO lines observed in J2310. In particular, our analysis suggests that the molecular gas of this source is characterized by a density ${\rm n\sim4\times10^{5} \ cm^{-3}}$, irradiated by a FUV flux ${\rm G_0\sim3\times10^3}$ (in Habing units of ${\rm 1.6\times10^{-3} \ erg \ s^{-1} \ cm^{-2}}$). No information on XDR can be inferred from our J2310 observations: this is expected, since the effect of X-ray photons on the molecular gas excitation can be disentangled from the one due to FUV photons only if CO lines with \jup$>10$ are available \citep[e.g][]{Schleicher:2010}.

\subsection{$\rm L_{FIR}$-normalised CO SLED}\label{J1148comparison}

Tight L$_{\rm CO}$-L$_{\rm FIR}$ correlations have been found over several orders of magnitude from low- to high-J CO lines \citep[$4<$~\jup~$~<13$;][]{Greve:2014, Liu:2015, Kamenetzky:2016, Lu:2017}  suggesting the average CO gas excitation conditions are mainly associated with star-formation activity, while AGN seem to have negligible impact on the CO SLED shape, at least for \jup~$\leq 10-13$. 
We therefore investigate the CO excitation conditions in our quasars by normalising their CO SLED to the FIR luminosity and by comparing the results with $z<1$ star-forming galaxies and AGN hosts \citep{Kamenetzky:2016}. 
The right panel of Fig.~\ref{fig:cosled} shows the FIR-normalised CO SLED  for J2310, J1319, J1148, the three $z\sim 6$ quasars by \cite{Venemans:2017a}, and for local starbursts and AGN host galaxies. 

We find that the L$_{\rm CO}$/L$_{\rm FIR}$ ratios of the three quasars are consistent, within the uncertainties, with the the FIR-normalised SLEDs observed locally, suggesting that the CO emission with \jup~$<7$ are powered by the star-formation activity in the host galaxies. A similar conclusion has been stated by \cite{Venemans:2017a} for their  three  quasars at $z\sim6$. By combining CO, [\ion{C}{ii}], and [\ion{C}{i}] observations, \cite{Venemans:2017a}  conclude that CO emission in their three quasars predominantly arises in PDR regions heated by young stars, and exclude substantial contribution from XDRs. 

From the right panel of Fig.~\ref{fig:cosled}, we also note that the CO/L$_{\rm FIR}$ ratios, up to \jup\ $=7$, of both J1319 and J1148 are below the  line ratios observed in J2310 and in the other high-$z$ quasars despite their comparable FIR luminosities. 
Such a discrepancy could indicate the heterogeneous properties  of PDRs regions among these high-$z$ quasars. Indeed the intensity of the CO line depends on the metallicity, gas temperature, molecular density, and radiation field strength \citep[e.g.][]{Uzgil:2016,Vallini:2018}.
In addition low CO-to-FIR line ratios can be also explained by an over-estimation of the FIR luminosity. Both AGN emission in the rest-frame mid-IR wavelengths (8~$\mu$m and 40~$\mu$m) and the presence of PAH features could affect the FIR luminosity estimates \citep[e.g.][]{Greve:2014}. 
In J1148 the AGN contamination at $\lambda_{\rm rest}>50$ \micron\ ($\nu_{\rm obs}\loa850$ GHz) is highly uncertain and may vary between 20\% and 70\% \citep{Leipski:2013,Schneider:2015}. The AGN contamination may explain the  magnitude of the CO/L$_{\rm FIR}$ discrepancy.  A deficiency in the molecular gas content could also explain the low CO-to-FIR line with respect to J2310.  As also observed in other lower-$z$ quasars characterised by extended ionised outflows \citep{Carniani:2015,Carniani:2017, Brusa:2018}, where a substantial gas fraction has been  is removed from the host galaxy by AGN winds. In the case of J1148 this scenario is supported by the detection of broad wings in the [\ion{C}{ii}] line profile, suggesting the presence of fast outflowing gas in the host galaxy \citep{Maiolino:2012,Cicone:2015}. In particular, \cite{Cicone:2015} found that J1148 hosts a powerful outflows with a mass outflow rate $>1000$ \sfr\ large enough to clear out all the gas content of the galaxy in less than $\sim6$ Myr. We further note that, for this quasar, \cite{Stefan:2015} infer a molecular gas mass from the CO(2-1) emission of $M_{\rm H2} = 2.6\times10^{10}$ \msun\ that is smaller than that measured in J2310 \citep[$M_{\rm H2} = 4.2\times10^{10}$ \msun][]{Shao:2019}. Given the  high dust mass of J1319 (Table~\ref{tab:prop}), we exclude that the outflow scenario would explain the low CO-to-FIR line ratios in this quasar.

\section{Discussion}\label{discussion}
\label{sec:discussion}
The nature of the discrepancy between the high CO(17-16) line luminosity observed in J1148 and local sources/high-$z$ quasars is unclear, both because of the complexity of the physical processes involved in the molecular gas excitation and the paucity of observational data. From the theoretical point of view, the emergence of bright high-J (\jup~$>$~13) CO lines can be associated both to extreme star formation, and AGN activity, and shocks induced by merging/outflows/SN-driven winds. In Table \ref{tab:tab3}, we report the properties of J2310, J1319, and J1148 that are more relevant in this context. 

\begin{itemize}

\item {\it Star formation:} The properties of J1148, J2310, and J1310 are quite similar in terms of FIR emission (Figure~\ref{fig:sed}). Assuming that dust heating is dominated by stars, the resulting SFRs among the three quasars are comparable (Table~\ref{tab:prop}). 
If high-J CO lines are predominantly excited by star-formation activity, we should have observed in J2310 and J1319 CO(17-16) lines as luminous as in J1148.

\item {\it AGN activity:} Strong high-$J$ CO lines can be excited in X-ray dominated regions \citep{Meijerink:2007,spaans:2008,vanderwerf:2010,Schleicher:2010}. Given the similarity among  black hole masses (M$_{\rm BH}~\sim 2-3\times10^{9}$~\msun) and AGN luminosities ($-27.80~<~{\rm M}_{1450}~<~-27.12$) of the three quasars, we do not expect their  X-ray luminosities to differ much.
However, we cannot exclude that the high luminosity of the CO(17-16) detected in J1148 can be associated with its X-ray radiation \citep{Gallerani:2014}. 
This interpretation is favoured by the fact that the X-ray luminosity ($L_{\rm X-ray}=1.4\times10^{45}$ erg/s) of J1148 is higher than that of J2310  ($L_{\rm X-ray}=7\times10^{44}$ erg/s; Fabio Vito, private communication, Vito et al. in prep.) suggesting the presence of a  stronger X-ray radiation in J1148.

\item {\it Shocks:} High temperatures associated with shock dominated regions can also be responsible for boosting the luminosity of high-$J$ CO lines \citep{panuzzo:2010,Hailey-Dunsheath:2012,Meijerink:2013}. In this context, it is remarkable that, while J1148 exhibits a massive, powerful outflow with a mass outflow rate $ > 1000$ \msun\ yr$^{-1}$ \citep{Maiolino:2012, Cicone:2015}, in the other sources no such strong outflows have been found (Carniani et al. in preparation).
 High resolution numerical simulations show that AGN feedback can trigger outflows as powerful as the one revealed in J1148 \citep{Barai:2018} and shock heat large quantity of gas to the high temperatures required for the excitation of high-J CO lines \citep{Costa:2014, Costa:2015, Costa:2018, Barai:2018}. Nevertheless, these simulations lack both the chemistry of molecular hydrogen \citep[e.g.,][]{Pallottini:2017a,Pallottini:2017b} and radiative transfer of X-ray photons \citep[e.g.,][]{Kakiichi:2017}, thus hampering a realistic comparison with observations.
\end{itemize}

To summarise, our analysis disfavours a scenario in which the high CO(17-16) luminosity observed in J1148 is driven by photodissociation regions, even accounting for the possible contamination by OH$^{+}$ emission. In the case of CO SLEDs excited by star formation, even in the case of high SFRs, the CO(6-5)- and FIR-normalised CO SLEDs are expected to decreases at high-J with a peak at $6<$~\jup~$<8$, as observed in the low-$z$ starburst sample and in J2310. Other mechanism associated with AGN (X-ray dominated regions, AGN-driven outflows) are more likely responsible for the excitation of high-J transitions.

However, the lack of CO(17-16) detections in J2310 and J1319 suggests that AGN activity is not always associated with strong high-J CO lines. 

\begin{table}
	\caption{Quasar properties}
	\label{tab:tab3}
	\begin{tabular}{lccc} 
		\hline
				 & J2310 & J1319 & J1148 \\
		\hline
		\hline
		M$_{1450}$$^{(1)}$ & -27.61 & -27.12 & -27.80 \\
		Log(SFR$_{\rm FIR}$/\sfr)$^{(2)}$ & $3.61\pm0.06$ & $3.4\pm0.3$ & $3.5\pm0.2$  \\
		L$_{\rm [\ion{C}{ii}]}$$^{(3)}$ [10$^{9}$ \lsun]& $8.7\pm1.4$ & $4.4\pm0.9$ & $37\pm9$ \\

		L$_{\rm X-ray}$$^{(4)}$ [10$^{45}~{\rm erg~s^{-1}}$]   & $0.7^{+0.5}_{-0.3}$ & - & $1.4^{+0.4}_{-0.3}$ \\	
		M$_{\rm BH}$$^{(5)}$ [10$^{9}$ \msun] & 1.8 & 2.7 & 3 \\
		\hline
		\hline
	\end{tabular}
	{\bf Note}: (1) Absolute magnitude at 1450\AA\ from \cite{Fan:2003}, \cite{Mortlock:2009} and \cite{Jiang:2016}; (2) SFR from FIR luminosity; (3) [\ion{C}{ii}] luminosities from \cite{Wang:2013} and \cite{Cicone:2015}; (4) X-ray luminosity between 2 keV and 10 keV from \cite{Gallerani:2017} and F. Vito, private communications (Vito et al. in prep.); (5) Black hole mass by \cite{Willott:2003}, \cite{Shao:2017}, and \cite{Feruglio:2018}.

\end{table}

\section{Conclusions}
\label{sec:conclusions}

We have presented ALMA observations of dust continuum and molecular CO line emission in two quasars at $z\sim6$, J1319 and J2310. We have also retrieved the CO and dust measurements for J1148 in order to compare the properties of these three quasars and understand the origin of their CO emission. Our main findings are:
\begin{enumerate}
\item  By fitting the continuum emission directly in the uv-plane, we have found that the bulk of dust emission arises from a compact ($<0.9$ kpc) region of the host galaxies. We have estimated a half-light radius of  ${\rm R_e = (0.66\pm0.05)~kpc}$ for J2310 and ${\rm R_e = (0.84\pm0.09)~kpc}$ for J1319.

\item We have performed a SED fitting on  our two quasars  as well as on J1148 using millimetre observations from literature and from this work and taking into account the impact of CMB on dust emission at  high redshift.  The dust properties, such as FIR luminosity, dust mass, temperature and emissivity index, of the J2310 and J1148 have been estimated with an uncertainties lower than 20\% (Table~\ref{tab:prop}).
The lack of high frequency ($\nu>400$~GHz) observations for J1319 have led to large uncertainties on dust temperature, which spans over a range $50~{\rm K}\lesssim T_{\rm dust}\lesssim 80~{\rm K}$. 
With a SFR$>1000$ \sfr\ and L$_{\rm FIR}>10^{13}$ \lsun, the host galaxies of the analysed quasars are in a starburst phase of their evolution.

\item In addition to the dust continuum emission we have discussed the observations of CO(9-8), CO(8-7), and CO(17-16) for J2310, in which we  have found a clear detection only for the first two lines. For what concerns J1319, we have presented ALMA observations of the CO(14-13), CO(17-16), and CO(19-18) lines and reported no detection in all the CO transitions observed. 

\item We have computed the CO SLEDs normalised to the CO(6-5) line and FIR luminosity for J2310, J1319, and J1148 and compared our results both with local starburst galaxies and AGN, and with other $z\sim 6$ quasars. 
For \jup$<9$, the CO(6-5)- and FIR-normalised CO SLEDs of $z\sim 6$ quasars are consistent with low-$z$ sources. However we note that the CO/FIR ratios of J1319 and J1148 are $\sim$2 times lower than those measured in J2310. This discrepancy could indicate heterogeneous ISM properties among high-$z$ quasars. We have also suggested that the difference of L$_{\rm CO}$/L$_{\rm FIR}$ between J2310 and J1148 can be explained by either a FIR AGN contamination or a lower molecular gas content in J1148. Latter scenario is supported by the presence of fast outflowing gas that can remove molecular gas from the host galaxies of these quasars.

\item The upper limits on the CO(17-16) transitions for J2310 and J1319  are consistent with those observed in local AGN and starburst galaxies; vice-versa the $3\sigma$ upper limits on the CO(17-16)/CO(6-5) ratio measured in these sources is 1$ \sigma$ lower than that measured in J1148. Mechanisms associated with AGN (X-ray dominated regions, AGN-driven outflows) are likely responsible for the excitation of this high-J CO transition and for the higher CO(17-16) luminosity measured in J1148. 

\end{enumerate}

In summary the no detection of high-J (\jup~$\geq 14$) CO transitions  in the quasars J2310 and J1319 reveals that AGN activity is not always associated with luminous ($\rm >10^8~L_{\odot}$) highly excited CO emission lines. The detection of \jup~$>8$ lines in $z\sim6$ quasars, complemented by X-ray observations and supported by dedicated high resolution numerical simulations (including AGN-driven feedback, molecular hydrogen chemistry, radiative transfer of X-ray photons), 
represent the best strategy to progress in the field and provide 
the optimal chances to understand the processes responsible for molecular gas excitation.

\section*{Acknowledgements}
This paper makes use of the following ALMA data: ADS/JAO.ALMA\#2013.1.00462.S, ADS/JAO.ALMA\#2015.1.01265.S, ADS/JAO.ALMA\#2012.1.00391.S. ALMA is a partnership of ESO (representing its member states), NSF (USA) and NINS (Japan), together with NRC (Canada), MOST and ASIAA (Taiwan), and KASI (Republic of Korea), in cooperation with the Republic of Chile. The Joint ALMA Observatory is operated by ESO, AUI/NRAO and NAOJ.
A.F. and S.C acknowledge support from the ERC Advanced Grant INTERSTELLAR H2020/740120. 
This work reflects only the authors' view and  the  European Research Commission is not responsible for information it contains. 
LV acknowledges funding from the European Union's Horizon 2020 research and innovation program under the Marie Sk\l{}odowska-Curie Grant agreement No. 746119. 
M.T. has been supported by the DISCSIM project, grant agreement 341137 funded by the European Research Council under ERC-2013-ADG.
RM acknowledges support from the ERC Advanced Grant 695671 `QUENCH' and from the Science and Technology Facilities Council (STFC).
CC and CF acknowledge support from the European Union Horizon 2020 research and innovation programme under the Marie Sk\l{}odowska-Curie grant agreement No 664931.

We are grateful to the anonymous referee for her/his comments. We finally acknowledge the contribution of Paolo Comaschi to the present work.
%




\bibliographystyle{mnras}
\bibliography{highJ_CO.bib} 

\begin{thebibliography}{}
\makeatletter
\relax
\def\mn@urlcharsother{\let\do\@makeother \do\$\do\&\do\#\do\^\do\_\do\%\do\~}
\def\mn@doi{\begingroup\mn@urlcharsother \@ifnextchar [ {\mn@doi@}
  {\mn@doi@[]}}
\def\mn@doi@[#1]#2{\def\@tempa{#1}\ifx\@tempa\@empty \href
  {http://dx.doi.org/#2} {doi:#2}\else \href {http://dx.doi.org/#2} {#1}\fi
  \endgroup}
\def\mn@eprint#1#2{\mn@eprint@#1:#2::\@nil}
\def\mn@eprint@arXiv#1{\href {http://arxiv.org/abs/#1} {{\tt arXiv:#1}}}
\def\mn@eprint@dblp#1{\href {http://dblp.uni-trier.de/rec/bibtex/#1.xml}
  {dblp:#1}}
\def\mn@eprint@#1:#2:#3:#4\@nil{\def\@tempa {#1}\def\@tempb {#2}\def\@tempc
  {#3}\ifx \@tempc \@empty \let \@tempc \@tempb \let \@tempb \@tempa \fi \ifx
  \@tempb \@empty \def\@tempb {arXiv}\fi \@ifundefined
  {mn@eprint@\@tempb}{\@tempb:\@tempc}{\expandafter \expandafter \csname
  mn@eprint@\@tempb\endcsname \expandafter{\@tempc}}}

\bibitem[\protect\citeauthoryear{{Alton}, {Xilouris}, {Misiriotis}, {Dasyra}
  \& {Dumke}}{{Alton} et~al.}{2004}]{Alton:2004}
{Alton} P.~B.,  {Xilouris} E.~M.,  {Misiriotis} A.,  {Dasyra} K.~M.,   {Dumke}
  M.,  2004, \mn@doi [\aap] {10.1051/0004-6361:20040438}, \href
  {http://adsabs.harvard.edu/abs/2004A%26A...425..109A} {425, 109}

\bibitem[\protect\citeauthoryear{{Aravena} et~al.,}{{Aravena}
  et~al.}{2016}]{Aravena:2016}
{Aravena} M.,  et~al., 2016, \mn@doi [\mnras] {10.1093/mnras/stw275}, \href
  {http://adsabs.harvard.edu/abs/2016MNRAS.457.4406A} {457, 4406}

\bibitem[\protect\citeauthoryear{{Ba{\~n}ados} et~al.,}{{Ba{\~n}ados}
  et~al.}{2016}]{Banados:2016}
{Ba{\~n}ados} E.,  et~al., 2016, \mn@doi [\apjs] {10.3847/0067-0049/227/1/11},
  \href {http://adsabs.harvard.edu/abs/2016ApJS..227...11B} {227, 11}

\bibitem[\protect\citeauthoryear{{Ba{\~n}ados} et~al.,}{{Ba{\~n}ados}
  et~al.}{2018}]{Banados:2018}
{Ba{\~n}ados} E.,  et~al., 2018, \mn@doi [\nat] {10.1038/nature25180}, \href
  {http://adsabs.harvard.edu/abs/2018Natur.553..473B} {553, 473}

\bibitem[\protect\citeauthoryear{{Barai}, {Gallerani}, {Pallottini}, {Ferrara},
  {Marconi}, {Cicone}, {Maiolino}  \& {Carniani}}{{Barai}
  et~al.}{2018}]{Barai:2018}
{Barai} P.,  {Gallerani} S.,  {Pallottini} A.,  {Ferrara} A.,  {Marconi} A.,
  {Cicone} C.,  {Maiolino} R.,   {Carniani} S.,  2018, \mn@doi [\mnras]
  {10.1093/mnras/stx2563}, \href
  {http://adsabs.harvard.edu/abs/2018MNRAS.473.4003B} {473, 4003}

\bibitem[\protect\citeauthoryear{{Beelen}, {Cox}, {Benford}, {Dowell},
  {Kov{\'a}cs}, {Bertoldi}, {Omont}  \& {Carilli}}{{Beelen}
  et~al.}{2006}]{Beelen:2006}
{Beelen} A.,  {Cox} P.,  {Benford} D.~J.,  {Dowell} C.~D.,  {Kov{\'a}cs} A.,
  {Bertoldi} F.,  {Omont} A.,   {Carilli} C.~L.,  2006, \mn@doi [\apj]
  {10.1086/500636}, \href {http://adsabs.harvard.edu/abs/2006ApJ...642..694B}
  {642, 694}

\bibitem[\protect\citeauthoryear{{Bertoldi}, {Carilli}, {Cox}, {Fan},
  {Strauss}, {Beelen}, {Omont}  \& {Zylka}}{{Bertoldi}
  et~al.}{2003a}]{Bertoldi:2003}
{Bertoldi} F.,  {Carilli} C.~L.,  {Cox} P.,  {Fan} X.,  {Strauss} M.~A.,
  {Beelen} A.,  {Omont} A.,   {Zylka} R.,  2003a, \mn@doi [\aap]
  {10.1051/0004-6361:20030710}, \href
  {http://adsabs.harvard.edu/abs/2003A%26A...406L..55B} {406, L55}

\bibitem[\protect\citeauthoryear{{Bertoldi} et~al.,}{{Bertoldi}
  et~al.}{2003b}]{Bertoldi:2003a}
{Bertoldi} F.,  et~al., 2003b, \mn@doi [\aap] {10.1051/0004-6361:20031345},
  \href {http://adsabs.harvard.edu/abs/2003A%26A...409L..47B} {409, L47}

\bibitem[\protect\citeauthoryear{{Bianchi} \& {Schneider}}{{Bianchi} \&
  {Schneider}}{2007}]{Bianchi:2007}
{Bianchi} S.,  {Schneider} R.,  2007, \mn@doi [\mnras]
  {10.1111/j.1365-2966.2007.11829.x}, \href
  {http://adsabs.harvard.edu/abs/2007MNRAS.378..973B} {378, 973}

\bibitem[\protect\citeauthoryear{{Bischetti}, {Maiolino}, {Fiore}, {Piconcelli}
   \& {Fluetsch}}{{Bischetti} et~al.}{2018}]{Bischetti:2018}
{Bischetti} M.,  {Maiolino} R.,  {Fiore} S. C.~F.,  {Piconcelli} E.,
  {Fluetsch} A.,  2018, arXiv e-prints, \href
  {https://ui.adsabs.harvard.edu/abs/2018arXiv180600786B} {p. arXiv:1806.00786}

\bibitem[\protect\citeauthoryear{{Brusa} et~al.,}{{Brusa}
  et~al.}{2018}]{Brusa:2018}
{Brusa} M.,  et~al., 2018, \mn@doi [\aap] {10.1051/0004-6361/201731641}, \href
  {http://adsabs.harvard.edu/abs/2018A%26A...612A..29B} {612, A29}

\bibitem[\protect\citeauthoryear{{Carilli} \& {Walter}}{{Carilli} \&
  {Walter}}{2013}]{Carilli:2013}
{Carilli} C.~L.,  {Walter} F.,  2013, \mn@doi [\araa]
  {10.1146/annurev-astro-082812-140953}, \href
  {http://adsabs.harvard.edu/abs/2013ARA%26A..51..105C} {51, 105}

\bibitem[\protect\citeauthoryear{{Carniani} et~al.,}{{Carniani}
  et~al.}{2013}]{Carniani:2013}
{Carniani} S.,  et~al., 2013, \mn@doi [\aap] {10.1051/0004-6361/201322320},
  \href {http://adsabs.harvard.edu/abs/2013A%26A...559A..29C} {559, A29}

\bibitem[\protect\citeauthoryear{{Carniani} et~al.,}{{Carniani}
  et~al.}{2015}]{Carniani:2015}
{Carniani} S.,  et~al., 2015, \mn@doi [\aap] {10.1051/0004-6361/201526557},
  \href {http://adsabs.harvard.edu/abs/2015A%26A...580A.102C} {580, A102}

\bibitem[\protect\citeauthoryear{{Carniani} et~al.,}{{Carniani}
  et~al.}{2017a}]{Carniani:2017a}
{Carniani} S.,  et~al., 2017a, \mn@doi [\aap] {10.1051/0004-6361/201630366},
  \href {http://adsabs.harvard.edu/abs/2017A%26A...605A..42C} {605, A42}

\bibitem[\protect\citeauthoryear{{Carniani} et~al.,}{{Carniani}
  et~al.}{2017b}]{Carniani:2017}
{Carniani} S.,  et~al., 2017b, \mn@doi [\aap] {10.1051/0004-6361/201730672},
  \href {http://adsabs.harvard.edu/abs/2017A%26A...605A.105C} {605, A105}

\bibitem[\protect\citeauthoryear{{Cicone} et~al.,}{{Cicone}
  et~al.}{2015}]{Cicone:2015}
{Cicone} C.,  et~al., 2015, \mn@doi [\aap] {10.1051/0004-6361/201424980}, \href
  {http://adsabs.harvard.edu/abs/2015A%26A...574A..14C} {574, A14}

\bibitem[\protect\citeauthoryear{{Costa}, {Sijacki}, {Trenti}  \&
  {Haehnelt}}{{Costa} et~al.}{2014}]{Costa:2014}
{Costa} T.,  {Sijacki} D.,  {Trenti} M.,   {Haehnelt} M.~G.,  2014, \mn@doi
  [\mnras] {10.1093/mnras/stu101}, \href
  {http://adsabs.harvard.edu/abs/2014MNRAS.439.2146C} {439, 2146}

\bibitem[\protect\citeauthoryear{{Costa}, {Sijacki}  \& {Haehnelt}}{{Costa}
  et~al.}{2015}]{Costa:2015}
{Costa} T.,  {Sijacki} D.,   {Haehnelt} M.~G.,  2015, \mn@doi [\mnras]
  {10.1093/mnrasl/slu193}, \href
  {http://adsabs.harvard.edu/abs/2015MNRAS.448L..30C} {448, L30}

\bibitem[\protect\citeauthoryear{{Costa}, {Rosdahl}, {Sijacki}  \&
  {Haehnelt}}{{Costa} et~al.}{2018}]{Costa:2018}
{Costa} T.,  {Rosdahl} J.,  {Sijacki} D.,   {Haehnelt} M.~G.,  2018, \mn@doi
  [\mnras] {10.1093/mnras/sty1514}, \href
  {http://adsabs.harvard.edu/abs/2018MNRAS.479.2079C} {479, 2079}

\bibitem[\protect\citeauthoryear{{De Rosa}, {Decarli}, {Walter}, {Fan},
  {Jiang}, {Kurk}, {Pasquali}  \& {Rix}}{{De Rosa} et~al.}{2011}]{De-Rosa:2011}
{De Rosa} G.,  {Decarli} R.,  {Walter} F.,  {Fan} X.,  {Jiang} L.,  {Kurk} J.,
  {Pasquali} A.,   {Rix} H.~W.,  2011, \mn@doi [\apj]
  {10.1088/0004-637X/739/2/56}, \href
  {http://adsabs.harvard.edu/abs/2011ApJ...739...56D} {739, 56}

\bibitem[\protect\citeauthoryear{{Decarli} et~al.,}{{Decarli}
  et~al.}{2017}]{Decarli:2017}
{Decarli} R.,  et~al., 2017, \mn@doi [\nat] {10.1038/nature22358}, \href
  {http://adsabs.harvard.edu/abs/2017Natur.545..457D} {545, 457}

\bibitem[\protect\citeauthoryear{{Decarli} et~al.,}{{Decarli}
  et~al.}{2018}]{Decarli:2018}
{Decarli} R.,  et~al., 2018, \mn@doi [\apj] {10.3847/1538-4357/aaa5aa}, \href
  {https://ui.adsabs.harvard.edu/abs/2018ApJ...854...97D} {854, 97}

\bibitem[\protect\citeauthoryear{{Di Matteo}, {Khandai}, {DeGraf}, {Feng},
  {Croft}, {Lopez}  \& {Springel}}{{Di Matteo} et~al.}{2012}]{Di-Matteo:2012}
{Di Matteo} T.,  {Khandai} N.,  {DeGraf} C.,  {Feng} Y.,  {Croft} R.~A.~C.,
  {Lopez} J.,   {Springel} V.,  2012, \mn@doi [\apjl]
  {10.1088/2041-8205/745/2/L29}, \href
  {http://adsabs.harvard.edu/abs/2012ApJ...745L..29D} {745, L29}

\bibitem[\protect\citeauthoryear{{Draine} \& {Lee}}{{Draine} \&
  {Lee}}{1984}]{Draine:1984}
{Draine} B.~T.,  {Lee} H.~M.,  1984, \mn@doi [\apj] {10.1086/162480}, \href
  {http://adsabs.harvard.edu/abs/1984ApJ...285...89D} {285, 89}

\bibitem[\protect\citeauthoryear{{Egami} et~al.,}{{Egami}
  et~al.}{2018}]{Egami:2018}
{Egami} E.,  et~al., 2018, \mn@doi [\pasa] {10.1017/pasa.2018.41}, \href
  {http://adsabs.harvard.edu/abs/2018PASA...35...48E} {35}

\bibitem[\protect\citeauthoryear{{Fan} et~al.,}{{Fan} et~al.}{2003}]{Fan:2003}
{Fan} X.,  et~al., 2003, \mn@doi [\aj] {10.1086/368246}, \href
  {http://adsabs.harvard.edu/abs/2003AJ....125.1649F} {125, 1649}

\bibitem[\protect\citeauthoryear{{Feruglio} et~al.,}{{Feruglio}
  et~al.}{2018}]{Feruglio:2018}
{Feruglio} C.,  et~al., 2018, \mn@doi [\aap] {10.1051/0004-6361/201833174},
  \href {http://adsabs.harvard.edu/abs/2018A%26A...619A..39F} {619, A39}

\bibitem[\protect\citeauthoryear{{Fixsen}, {Cheng}, {Gales}, {Mather}, {Shafer}
   \& {Wright}}{{Fixsen} et~al.}{1996}]{Fixsen:1996}
{Fixsen} D.~J.,  {Cheng} E.~S.,  {Gales} J.~M.,  {Mather} J.~C.,  {Shafer}
  R.~A.,   {Wright} E.~L.,  1996, \mn@doi [\apj] {10.1086/178173}, \href
  {http://adsabs.harvard.edu/abs/1996ApJ...473..576F} {473, 576}

\bibitem[\protect\citeauthoryear{{Foreman-Mackey}, {Hogg}, {Lang}  \&
  {Goodman}}{{Foreman-Mackey} et~al.}{2013}]{Foreman-Mackey:2013}
{Foreman-Mackey} D.,  {Hogg} D.~W.,  {Lang} D.,   {Goodman} J.,  2013, \mn@doi
  [\pasp] {10.1086/670067}, \href
  {http://adsabs.harvard.edu/abs/2013PASP..125..306F} {125, 306}

\bibitem[\protect\citeauthoryear{{Gallerani} et~al.,}{{Gallerani}
  et~al.}{2012}]{Gallerani:2012}
{Gallerani} S.,  et~al., 2012, \mn@doi [\aap] {10.1051/0004-6361/201118705},
  \href {http://adsabs.harvard.edu/abs/2012A%26A...543A.114G} {543, A114}

\bibitem[\protect\citeauthoryear{{Gallerani}, {Ferrara}, {Neri}  \&
  {Maiolino}}{{Gallerani} et~al.}{2014}]{Gallerani:2014}
{Gallerani} S.,  {Ferrara} A.,  {Neri} R.,   {Maiolino} R.,  2014, \mn@doi
  [\mnras] {10.1093/mnras/stu2031}, \href
  {http://adsabs.harvard.edu/abs/2014MNRAS.445.2848G} {445, 2848}

\bibitem[\protect\citeauthoryear{{Gallerani}, {Fan}, {Maiolino}  \&
  {Pacucci}}{{Gallerani} et~al.}{2017a}]{Gallerani:2017a}
{Gallerani} S.,  {Fan} X.,  {Maiolino} R.,   {Pacucci} F.,  2017a, \mn@doi
  [\pasa] {10.1017/pasa.2017.14}, \href
  {http://adsabs.harvard.edu/abs/2017PASA...34...22G} {34, e022}

\bibitem[\protect\citeauthoryear{{Gallerani} et~al.,}{{Gallerani}
  et~al.}{2017b}]{Gallerani:2017}
{Gallerani} S.,  et~al., 2017b, \mn@doi [\mnras] {10.1093/mnras/stx363}, \href
  {http://adsabs.harvard.edu/abs/2017MNRAS.467.3590G} {467, 3590}

\bibitem[\protect\citeauthoryear{{Genzel} et~al.,}{{Genzel}
  et~al.}{2015}]{Genzel:2015}
{Genzel} R.,  et~al., 2015, \mn@doi [\apj] {10.1088/0004-637X/800/1/20}, \href
  {http://adsabs.harvard.edu/abs/2015ApJ...800...20G} {800, 20}

\bibitem[\protect\citeauthoryear{{Giallongo}, {Grazian}, {Fiore}, {Fontana}  \&
  et al.}{{Giallongo} et~al.}{2015}]{Giallongo:2015a}
{Giallongo} E.,  {Grazian} A.,  {Fiore} F.,  {Fontana} A.,   et al. 2015,
  \mn@doi [\aap] {10.1051/0004-6361/201425334}, \href
  {http://adsabs.harvard.edu/abs/2015A%26A...578A..83G} {578, A83}

\bibitem[\protect\citeauthoryear{{Gonz{\'a}lez-Alfonso}
  et~al.,}{{Gonz{\'a}lez-Alfonso} et~al.}{2013}]{Gonzalez-Alfonso:2013}
{Gonz{\'a}lez-Alfonso} E.,  et~al., 2013, \mn@doi [\aap]
  {10.1051/0004-6361/201220466}, \href
  {https://ui.adsabs.harvard.edu/\#abs/2013A&A...550A..25G} {550, A25}

\bibitem[\protect\citeauthoryear{{Gonz{\'a}lez-Alfonso}
  et~al.,}{{Gonz{\'a}lez-Alfonso} et~al.}{2018}]{Gonzalez-Alfonso:2018}
{Gonz{\'a}lez-Alfonso} E.,  et~al., 2018, \mn@doi [\apj]
  {10.3847/1538-4357/aab6b8}, \href
  {http://adsabs.harvard.edu/abs/2018ApJ...857...66G} {857, 66}

\bibitem[\protect\citeauthoryear{{Greve} et~al.,}{{Greve}
  et~al.}{2014}]{Greve:2014}
{Greve} T.~R.,  et~al., 2014, \mn@doi [\apj] {10.1088/0004-637X/794/2/142},
  \href {http://adsabs.harvard.edu/abs/2014ApJ...794..142G} {794, 142}

\bibitem[\protect\citeauthoryear{{Hailey-Dunsheath} et~al.,}{{Hailey-Dunsheath}
  et~al.}{2012}]{Hailey-Dunsheath:2012}
{Hailey-Dunsheath} S.,  et~al., 2012, \mn@doi [\apj]
  {10.1088/0004-637X/755/1/57}, \href
  {http://adsabs.harvard.edu/abs/2012ApJ...755...57H} {755, 57}

\bibitem[\protect\citeauthoryear{{Hashimoto}, {Inoue}, {Tamura}, {Matsuo},
  {Mawatari}  \& {Yamaguchi}}{{Hashimoto} et~al.}{2018}]{Hashimoto:2018a}
{Hashimoto} T.,  {Inoue} A.~K.,  {Tamura} Y.,  {Matsuo} H.,  {Mawatari} K.,
  {Yamaguchi} Y.,  2018, preprint, \href
  {http://adsabs.harvard.edu/abs/2018arXiv181100030H} {} (\mn@eprint {arXiv}
  {1811.00030})

\bibitem[\protect\citeauthoryear{{Jiang}, {Fan}, {Vestergaard}, {Kurk},
  {Walter}, {Kelly}  \& {Strauss}}{{Jiang} et~al.}{2007}]{Jiang:2007}
{Jiang} L.,  {Fan} X.,  {Vestergaard} M.,  {Kurk} J.~D.,  {Walter} F.,  {Kelly}
  B.~C.,   {Strauss} M.~A.,  2007, \mn@doi [\aj] {10.1086/520811}, \href
  {http://adsabs.harvard.edu/abs/2007AJ....134.1150J} {134, 1150}

\bibitem[\protect\citeauthoryear{{Jiang}, {McGreer}, {Fan}, {Strauss},
  {Ba{\~n}ados}, {Becker}, {Bian}  \& {Farnsworth}}{{Jiang}
  et~al.}{2016}]{Jiang:2016}
{Jiang} L.,  {McGreer} I.~D.,  {Fan} X.,  {Strauss} M.~A.,  {Ba{\~n}ados} E.,
  {Becker} R.~H.,  {Bian} F.,   {Farnsworth} K.,  2016, \mn@doi [\apj]
  {10.3847/1538-4357/833/2/222}, \href
  {http://adsabs.harvard.edu/abs/2016ApJ...833..222J} {833, 222}

\bibitem[\protect\citeauthoryear{{Kakiichi}, {Graziani}, {Ciardi}, {Meiksin},
  {Compostella}, {Eide}  \& {Zaroubi}}{{Kakiichi} et~al.}{2017}]{Kakiichi:2017}
{Kakiichi} K.,  {Graziani} L.,  {Ciardi} B.,  {Meiksin} A.,  {Compostella} M.,
  {Eide} M.~B.,   {Zaroubi} S.,  2017, \mn@doi [\mnras] {10.1093/mnras/stx603},
  \href {http://adsabs.harvard.edu/abs/2017MNRAS.468.3718K} {468, 3718}

\bibitem[\protect\citeauthoryear{{Kakkad} et~al.,}{{Kakkad}
  et~al.}{2017}]{Kakkad:2017}
{Kakkad} D.,  et~al., 2017, \mn@doi [\mnras] {10.1093/mnras/stx726}, \href
  {http://adsabs.harvard.edu/abs/2017MNRAS.468.4205K} {468, 4205}

\bibitem[\protect\citeauthoryear{{Kamenetzky}, {Rangwala}, {Glenn}, {Maloney}
  \& {Conley}}{{Kamenetzky} et~al.}{2016}]{Kamenetzky:2016}
{Kamenetzky} J.,  {Rangwala} N.,  {Glenn} J.,  {Maloney} P.~R.,   {Conley} A.,
  2016, \mn@doi [\apj] {10.3847/0004-637X/829/2/93}, \href
  {https://ui.adsabs.harvard.edu/\#abs/2016ApJ...829...93K} {829, 93}

\bibitem[\protect\citeauthoryear{{Kennicutt} \& {Evans}}{{Kennicutt} \&
  {Evans}}{2012}]{Kennicutt:2012}
{Kennicutt} R.~C.,  {Evans} N.~J.,  2012, \mn@doi [\araa]
  {10.1146/annurev-astro-081811-125610}, \href
  {http://adsabs.harvard.edu/abs/2012ARA%26A..50..531K} {50, 531}

\bibitem[\protect\citeauthoryear{{Kormendy} \& {Ho}}{{Kormendy} \&
  {Ho}}{2013}]{Kormendy:2013}
{Kormendy} J.,  {Ho} L.~C.,  2013, \mn@doi [\araa]
  {10.1146/annurev-astro-082708-101811}, \href
  {http://adsabs.harvard.edu/abs/2013ARA%26A..51..511K} {51, 511}

\bibitem[\protect\citeauthoryear{{Kulkarni}, {Worseck}  \&
  {Hennawi}}{{Kulkarni} et~al.}{2019}]{Kulkarni:2018}
{Kulkarni} G.,  {Worseck} G.,   {Hennawi} J.~F.,  2019, \mn@doi [Monthly
  Notices of the Royal Astronomical Society] {10.1093/mnras/stz1493}, \href
  {https://ui.adsabs.harvard.edu/abs/2019MNRAS.488.1035K} {488, 1035}

\bibitem[\protect\citeauthoryear{{Kurk}, {Walter}, {Fan}, {Jiang}, {Riechers},
  {Rix}, {Pentericci}  \& {Strauss}}{{Kurk} et~al.}{2007}]{Kurk:2007}
{Kurk} J.~D.,  {Walter} F.,  {Fan} X.,  {Jiang} L.,  {Riechers} D.~A.,  {Rix}
  H.-W.,  {Pentericci} L.,   {Strauss} 2007, \mn@doi [\apj] {10.1086/521596},
  \href {http://adsabs.harvard.edu/abs/2007ApJ...669...32K} {669, 32}

\bibitem[\protect\citeauthoryear{{Lamastra}, {Menci}, {Maiolino}, {Fiore}  \&
  {Merloni}}{{Lamastra} et~al.}{2010}]{Lamastra:2010}
{Lamastra} A.,  {Menci} N.,  {Maiolino} R.,  {Fiore} F.,   {Merloni} A.,  2010,
  \mn@doi [\mnras] {10.1111/j.1365-2966.2010.16439.x}, \href
  {http://adsabs.harvard.edu/abs/2010MNRAS.405...29L} {405, 29}

\bibitem[\protect\citeauthoryear{{Leipski} et~al.,}{{Leipski}
  et~al.}{2013}]{Leipski:2013}
{Leipski} C.,  et~al., 2013, \mn@doi [\apj] {10.1088/0004-637X/772/2/103},
  \href {http://adsabs.harvard.edu/abs/2013ApJ...772..103L} {772, 103}

\bibitem[\protect\citeauthoryear{{Li} et~al.,}{{Li} et~al.}{2007}]{Li:2007}
{Li} Y.,  et~al., 2007, \mn@doi [\apj] {10.1086/519297}, \href
  {http://adsabs.harvard.edu/abs/2007ApJ...665..187L} {665, 187}

\bibitem[\protect\citeauthoryear{{Liu}, {Gao}, {Isaak}, {Daddi}, {Yang}, {Lu}
  \& {van der Werf}}{{Liu} et~al.}{2015}]{Liu:2015}
{Liu} D.,  {Gao} Y.,  {Isaak} K.,  {Daddi} E.,  {Yang} C.,  {Lu} N.,   {van der
  Werf} P.,  2015, \mn@doi [\apjl] {10.1088/2041-8205/810/2/L14}, \href
  {http://adsabs.harvard.edu/abs/2015ApJ...810L..14L} {810, L14}

\bibitem[\protect\citeauthoryear{{Lu} et~al.,}{{Lu} et~al.}{2017}]{Lu:2017}
{Lu} N.,  et~al., 2017, \mn@doi [The Astrophysical Journal Supplement Series]
  {10.3847/1538-4365/aa6476}, \href
  {https://ui.adsabs.harvard.edu/\#abs/2017ApJS..230....1L} {230, 1}

\bibitem[\protect\citeauthoryear{{Madau} \& {Haardt}}{{Madau} \&
  {Haardt}}{2015}]{Madau:2015a}
{Madau} P.,  {Haardt} F.,  2015, \mn@doi [\apjl] {10.1088/2041-8205/813/1/L8},
  \href {http://adsabs.harvard.edu/abs/2015ApJ...813L...8M} {813, L8}

\bibitem[\protect\citeauthoryear{{Maiolino} et~al.,}{{Maiolino}
  et~al.}{2012}]{Maiolino:2012}
{Maiolino} R.,  et~al., 2012, \mn@doi [\mnras]
  {10.1111/j.1745-3933.2012.01303.x}, \href
  {http://adsabs.harvard.edu/abs/2012MNRAS.425L..66M} {425, L66}

\bibitem[\protect\citeauthoryear{{Manti}, {Gallerani}, {Ferrara}, {Greig}  \&
  {Feruglio}}{{Manti} et~al.}{2017}]{Manti:2017}
{Manti} S.,  {Gallerani} S.,  {Ferrara} A.,  {Greig} B.,   {Feruglio} C.,
  2017, \mn@doi [\mnras] {10.1093/mnras/stw3168}, \href
  {http://adsabs.harvard.edu/abs/2017MNRAS.466.1160M} {466, 1160}

\bibitem[\protect\citeauthoryear{{Mashian}, {Sturm}, {Sternberg}, {Janssen},
  {Hailey-Dunsheath}, {Fischer}, {Contursi}  \&
  {Gonz{\'a}lez-Alfonso}}{{Mashian} et~al.}{2015}]{Mashian:2015}
{Mashian} N.,  {Sturm} E.,  {Sternberg} A.,  {Janssen} A.,  {Hailey-Dunsheath}
  S.,  {Fischer} J.,  {Contursi} A.,   {Gonz{\'a}lez-Alfonso} E.,  2015,
  \mn@doi [\apj] {10.1088/0004-637X/802/2/81}, \href
  {http://adsabs.harvard.edu/abs/2015ApJ...802...81M} {802, 81}

\bibitem[\protect\citeauthoryear{{Matsuoka} et~al.,}{{Matsuoka}
  et~al.}{2016}]{Matsuoka:2016}
{Matsuoka} Y.,  et~al., 2016, \mn@doi [\apj] {10.3847/0004-637X/828/1/26},
  \href {http://adsabs.harvard.edu/abs/2016ApJ...828...26M} {828, 26}

\bibitem[\protect\citeauthoryear{{Matsuoka} et~al.,}{{Matsuoka}
  et~al.}{2018}]{Matsuoka:2018}
{Matsuoka} Y.,  et~al., 2018, \mn@doi [\pasj] {10.1093/pasj/psx046}, \href
  {http://adsabs.harvard.edu/abs/2018PASJ...70S..35M} {70, S35}

\bibitem[\protect\citeauthoryear{{Mazzucchelli} et~al.,}{{Mazzucchelli}
  et~al.}{2017}]{Mazzucchelli:2017}
{Mazzucchelli} C.,  et~al., 2017, \mn@doi [\apj] {10.3847/1538-4357/aa9185},
  \href {http://adsabs.harvard.edu/abs/2017ApJ...849...91M} {849, 91}

\bibitem[\protect\citeauthoryear{{McMullin}, {Waters}, {Schiebel}, {Young}  \&
  {Golap}}{{McMullin} et~al.}{2007}]{McMullin:2007}
{McMullin} J.~P.,  {Waters} B.,  {Schiebel} D.,  {Young} W.,   {Golap} K.,
  2007, in {Shaw} R.~A.,  {Hill} F.,   {Bell} D.~J.,  eds,  Astronomical
  Society of the Pacific Conference Series Vol. 376, Astronomical Data Analysis
  Software and Systems XVI. p.~127

\bibitem[\protect\citeauthoryear{{Meijerink} \& {Spaans}}{{Meijerink} \&
  {Spaans}}{2005}]{Meijerink:2005}
{Meijerink} R.,  {Spaans} M.,  2005, \mn@doi [\aap]
  {10.1051/0004-6361:20042398}, \href
  {http://adsabs.harvard.edu/abs/2005A%26A...436..397M} {436, 397}

\bibitem[\protect\citeauthoryear{{Meijerink}, {Spaans}  \&
  {Israel}}{{Meijerink} et~al.}{2007}]{Meijerink:2007}
{Meijerink} R.,  {Spaans} M.,   {Israel} F.~P.,  2007, \mn@doi [\aap]
  {10.1051/0004-6361:20066130}, \href
  {http://adsabs.harvard.edu/abs/2007A%26A...461..793M} {461, 793}

\bibitem[\protect\citeauthoryear{{Meijerink} et~al.,}{{Meijerink}
  et~al.}{2013}]{Meijerink:2013}
{Meijerink} R.,  et~al., 2013, \mn@doi [\apjl] {10.1088/2041-8205/762/2/L16},
  \href {http://adsabs.harvard.edu/abs/2013ApJ...762L..16M} {762, L16}

\bibitem[\protect\citeauthoryear{{Mingozzi} et~al.,}{{Mingozzi}
  et~al.}{2018}]{Mingozzi:2018}
{Mingozzi} M.,  et~al., 2018, \mn@doi [\mnras] {10.1093/mnras/stx3011}, \href
  {https://ui.adsabs.harvard.edu/\#abs/2018MNRAS.474.3640M} {474, 3640}

\bibitem[\protect\citeauthoryear{{Mitra}, {Choudhury}  \& {Ferrara}}{{Mitra}
  et~al.}{2018}]{Mitra:2018}
{Mitra} S.,  {Choudhury} T.~R.,   {Ferrara} A.,  2018, \mn@doi [\mnras]
  {10.1093/mnras/stx2443}, \href
  {http://adsabs.harvard.edu/abs/2018MNRAS.473.1416M} {473, 1416}

\bibitem[\protect\citeauthoryear{{Mortlock}, {Patel}, {Warren}, {Venemans},
  {McMahon}, {Hewett}, {Simpson}  \& {Sharp}}{{Mortlock}
  et~al.}{2009}]{Mortlock:2009}
{Mortlock} D.~J.,  {Patel} M.,  {Warren} S.~J.,  {Venemans} B.~P.,  {McMahon}
  R.~G.,  {Hewett} P.~C.,  {Simpson} C.,   {Sharp} R.~G.,  2009, \mn@doi [\aap]
  {10.1051/0004-6361/200811161}, \href
  {http://adsabs.harvard.edu/abs/2009A%26A...505...97M} {505, 97}

\bibitem[\protect\citeauthoryear{{Narayanan} et~al.,}{{Narayanan}
  et~al.}{2008}]{Narayanan:2008}
{Narayanan} D.,  et~al., 2008, \mn@doi [\apjs] {10.1086/521776}, \href
  {http://adsabs.harvard.edu/abs/2008ApJS..174...13N} {174, 13}

\bibitem[\protect\citeauthoryear{{Obreschkow}, {Heywood}, {Kl{\"o}ckner}  \&
  {Rawlings}}{{Obreschkow} et~al.}{2009}]{Obreschkow:2009}
{Obreschkow} D.,  {Heywood} I.,  {Kl{\"o}ckner} H.-R.,   {Rawlings} S.,  2009,
  \mn@doi [\apj] {10.1088/0004-637X/702/2/1321}, \href
  {http://adsabs.harvard.edu/abs/2009ApJ...702.1321O} {702, 1321}

\bibitem[\protect\citeauthoryear{{Pallottini}, {Ferrara}, {Gallerani},
  {Vallini}, {Maiolino}  \& {Salvadori}}{{Pallottini}
  et~al.}{2017a}]{Pallottini:2017a}
{Pallottini} A.,  {Ferrara} A.,  {Gallerani} S.,  {Vallini} L.,  {Maiolino} R.,
    {Salvadori} S.,  2017a, \mn@doi [\mnras] {10.1093/mnras/stw2847}, \href
  {http://adsabs.harvard.edu/abs/2017MNRAS.465.2540P} {465, 2540}

\bibitem[\protect\citeauthoryear{{Pallottini}, {Ferrara}, {Bovino}, {Vallini},
  {Gallerani}, {Maiolino}  \& {Salvadori}}{{Pallottini}
  et~al.}{2017b}]{Pallottini:2017b}
{Pallottini} A.,  {Ferrara} A.,  {Bovino} S.,  {Vallini} L.,  {Gallerani} S.,
  {Maiolino} R.,   {Salvadori} S.,  2017b, \mn@doi [\mnras]
  {10.1093/mnras/stx1792}, \href
  {http://adsabs.harvard.edu/abs/2017MNRAS.471.4128P} {471, 4128}

\bibitem[\protect\citeauthoryear{{Panuzzo} et~al.,}{{Panuzzo}
  et~al.}{2010}]{panuzzo:2010}
{Panuzzo} P.,  et~al., 2010, \mn@doi [\aap] {10.1051/0004-6361/201014558},
  \href {http://adsabs.harvard.edu/abs/2010A%26A...518L..37P} {518, L37}

\bibitem[\protect\citeauthoryear{{Parsa}, {Dunlop}  \& {McLure}}{{Parsa}
  et~al.}{2018}]{Parsa:2018}
{Parsa} S.,  {Dunlop} J.~S.,   {McLure} R.~J.,  2018, \mn@doi [\mnras]
  {10.1093/mnras/stx2887}, \href
  {http://adsabs.harvard.edu/abs/2018MNRAS.474.2904P} {474, 2904}

\bibitem[\protect\citeauthoryear{{Pilbratt} et~al.,}{{Pilbratt}
  et~al.}{2010}]{Pilbratt:2010}
{Pilbratt} G.~L.,  et~al., 2010, \mn@doi [\aap] {10.1051/0004-6361/201014759},
  \href {http://adsabs.harvard.edu/abs/2010A%26A...518L...1P} {518, L1}

\bibitem[\protect\citeauthoryear{{Planck Collaboration} et~al.,}{{Planck
  Collaboration} et~al.}{2016}]{Planck-Collaboration:2015}
{Planck Collaboration} et~al., 2016, \mn@doi [Astronomy and Astrophysics]
  {10.1051/0004-6361/201525830}, \href
  {https://ui.adsabs.harvard.edu/abs/2016A&A...594A..13P} {594, A13}

\bibitem[\protect\citeauthoryear{{Poglitsch} et~al.,}{{Poglitsch}
  et~al.}{2010}]{Poglitsch:2010}
{Poglitsch} A.,  et~al., 2010, \mn@doi [\aap] {10.1051/0004-6361/201014535},
  \href {http://adsabs.harvard.edu/abs/2010A%26A...518L...2P} {518, L2}

\bibitem[\protect\citeauthoryear{{Qin} et~al.,}{{Qin} et~al.}{2017}]{Qin:2017}
{Qin} Y.,  et~al., 2017, \mn@doi [\mnras] {10.1093/mnras/stx1909}, \href
  {http://adsabs.harvard.edu/abs/2017MNRAS.472.2009Q} {472, 2009}

\bibitem[\protect\citeauthoryear{{Richings} \&
  {Faucher-Gigu{\`e}re}}{{Richings} \&
  {Faucher-Gigu{\`e}re}}{2018}]{Richings:2018}
{Richings} A.~J.,  {Faucher-Gigu{\`e}re} C.-A.,  2018, \mn@doi [\mnras]
  {10.1093/mnras/stx3014}, \href
  {https://ui.adsabs.harvard.edu/\#abs/2018MNRAS.474.3673R} {474, 3673}

\bibitem[\protect\citeauthoryear{{Riechers} et~al.,}{{Riechers}
  et~al.}{2009}]{Riechers:2009}
{Riechers} D.~A.,  et~al., 2009, \mn@doi [\apj] {10.1088/0004-637X/703/2/1338},
  \href {http://adsabs.harvard.edu/abs/2009ApJ...703.1338R} {703, 1338}

\bibitem[\protect\citeauthoryear{{Riechers} et~al.,}{{Riechers}
  et~al.}{2013}]{Riechers:2013}
{Riechers} D.~A.,  et~al., 2013, \mn@doi [\nat] {10.1038/nature12050}, \href
  {http://adsabs.harvard.edu/abs/2013Natur.496..329R} {496, 329}

\bibitem[\protect\citeauthoryear{{Rosenberg} et~al.,}{{Rosenberg}
  et~al.}{2015}]{Rosenberg:2015}
{Rosenberg} M.~J.~F.,  et~al., 2015, \mn@doi [\apj]
  {10.1088/0004-637X/801/2/72}, \href
  {https://ui.adsabs.harvard.edu/abs/2015ApJ...801...72R} {801, 72}

\bibitem[\protect\citeauthoryear{{Schleicher}, {Spaans}  \&
  {Klessen}}{{Schleicher} et~al.}{2010}]{Schleicher:2010}
{Schleicher} D.~R.~G.,  {Spaans} M.,   {Klessen} R.~S.,  2010, \mn@doi [\aap]
  {10.1051/0004-6361/200913467}, \href
  {http://adsabs.harvard.edu/abs/2010A%26A...513A...7S} {513, A7}

\bibitem[\protect\citeauthoryear{{Schneider}, {Bianchi}, {Valiante}, {Risaliti}
   \& {Salvadori}}{{Schneider} et~al.}{2015}]{Schneider:2015}
{Schneider} R.,  {Bianchi} S.,  {Valiante} R.,  {Risaliti} G.,   {Salvadori}
  S.,  2015, \mn@doi [\aap] {10.1051/0004-6361/201526105}, \href
  {https://ui.adsabs.harvard.edu/abs/2015A&A...579A..60S} {579, A60}

\bibitem[\protect\citeauthoryear{{Shao} et~al.,}{{Shao}
  et~al.}{2017}]{Shao:2017}
{Shao} Y.,  et~al., 2017, \mn@doi [\apj] {10.3847/1538-4357/aa826c}, \href
  {http://adsabs.harvard.edu/abs/2017ApJ...845..138S} {845, 138}

\bibitem[\protect\citeauthoryear{{Shao} et~al.,}{{Shao}
  et~al.}{2019}]{Shao:2019}
{Shao} Y.,  et~al., 2019, \mn@doi [The Astrophysical Journal]
  {10.3847/1538-4357/ab133d}, \href
  {https://ui.adsabs.harvard.edu/abs/2019ApJ...876...99S} {876, 99}

\bibitem[\protect\citeauthoryear{{Spaans} \& {Meijerink}}{{Spaans} \&
  {Meijerink}}{2008}]{spaans:2008}
{Spaans} M.,  {Meijerink} R.,  2008, \mn@doi [\apjl] {10.1086/588253}, \href
  {http://adsabs.harvard.edu/abs/2008ApJ...678L...5S} {678, L5}

\bibitem[\protect\citeauthoryear{{Spinoglio} et~al.,}{{Spinoglio}
  et~al.}{2017}]{Spinoglio:2017}
{Spinoglio} L.,  et~al., 2017, \mn@doi [\pasa] {10.1017/pasa.2017.48}, \href
  {http://adsabs.harvard.edu/abs/2017PASA...34...57S} {34, e057}

\bibitem[\protect\citeauthoryear{{Stefan} et~al.,}{{Stefan}
  et~al.}{2015}]{Stefan:2015}
{Stefan} I.~I.,  et~al., 2015, \mn@doi [\mnras] {10.1093/mnras/stv1108}, \href
  {http://adsabs.harvard.edu/abs/2015MNRAS.451.1713S} {451, 1713}

\bibitem[\protect\citeauthoryear{{Tazzari}, {Beaujean}  \& {Testi}}{{Tazzari}
  et~al.}{2018}]{Tazzari:2018}
{Tazzari} M.,  {Beaujean} F.,   {Testi} L.,  2018, \mn@doi [\mnras]
  {10.1093/mnras/sty409}, \href
  {http://adsabs.harvard.edu/abs/2018MNRAS.476.4527T} {476, 4527}

\bibitem[\protect\citeauthoryear{{Tunnard} \& {Greve}}{{Tunnard} \&
  {Greve}}{2016}]{Tunnard:2016}
{Tunnard} R.,  {Greve} T.~R.,  2016, \mn@doi [\apj]
  {10.3847/0004-637X/819/2/161}, \href
  {http://adsabs.harvard.edu/abs/2016ApJ...819..161T} {819, 161}

\bibitem[\protect\citeauthoryear{{Uzgil}, {Bradford}, {Hailey-Dunsheath},
  {Maloney}  \& {Aguirre}}{{Uzgil} et~al.}{2016}]{Uzgil:2016}
{Uzgil} B.~D.,  {Bradford} C.~M.,  {Hailey-Dunsheath} S.,  {Maloney} P.~R.,
  {Aguirre} J.~E.,  2016, \mn@doi [The Astrophysical Journal]
  {10.3847/0004-637X/832/2/209}, \href
  {https://ui.adsabs.harvard.edu/abs/2016ApJ...832..209U} {832, 209}

\bibitem[\protect\citeauthoryear{{Valiante}, {Schneider}, {Salvadori}  \&
  {Gallerani}}{{Valiante} et~al.}{2014}]{Valiante:2014}
{Valiante} R.,  {Schneider} R.,  {Salvadori} S.,   {Gallerani} S.,  2014,
  \mn@doi [\mnras] {10.1093/mnras/stu1613}, \href
  {http://adsabs.harvard.edu/abs/2014MNRAS.444.2442V} {444, 2442}

\bibitem[\protect\citeauthoryear{{Valiante}, {Agarwal}, {Habouzit}  \&
  {Pezzulli}}{{Valiante} et~al.}{2017}]{Valiante:2017}
{Valiante} R.,  {Agarwal} B.,  {Habouzit} M.,   {Pezzulli} E.,  2017, \mn@doi
  [\pasa] {10.1017/pasa.2017.25}, \href
  {http://adsabs.harvard.edu/abs/2017PASA...34...31V} {34, e031}

\bibitem[\protect\citeauthoryear{{Vallini}, {Pallottini}, {Ferrara},
  {Gallerani}, {Sobacchi}  \& {Behrens}}{{Vallini} et~al.}{2018}]{Vallini:2018}
{Vallini} L.,  {Pallottini} A.,  {Ferrara} A.,  {Gallerani} S.,  {Sobacchi} E.,
    {Behrens} C.,  2018, \mn@doi [\mnras] {10.1093/mnras/stx2376}, \href
  {http://adsabs.harvard.edu/abs/2018MNRAS.473..271V} {473, 271}

\bibitem[\protect\citeauthoryear{{Venemans} et~al.,}{{Venemans}
  et~al.}{2017a}]{Venemans:2017}
{Venemans} B.~P.,  et~al., 2017a, \mn@doi [\apj] {10.3847/1538-4357/aa62ac},
  \href {http://adsabs.harvard.edu/abs/2017ApJ...837..146V} {837, 146}

\bibitem[\protect\citeauthoryear{{Venemans} et~al.,}{{Venemans}
  et~al.}{2017b}]{Venemans:2017a}
{Venemans} B.~P.,  et~al., 2017b, \mn@doi [\apj] {10.3847/1538-4357/aa81cb},
  \href {http://adsabs.harvard.edu/abs/2017ApJ...845..154V} {845, 154}

\bibitem[\protect\citeauthoryear{{Volonteri}}{{Volonteri}}{2010}]{Volonteri:2010}
{Volonteri} M.,  2010, \mn@doi [\aapr] {10.1007/s00159-010-0029-x}, \href
  {http://adsabs.harvard.edu/abs/2010A%26ARv..18..279V} {18, 279}

\bibitem[\protect\citeauthoryear{{Volonteri} \& {Gnedin}}{{Volonteri} \&
  {Gnedin}}{2009}]{Volonteri:2009}
{Volonteri} M.,  {Gnedin} N.~Y.,  2009, \mn@doi [\apj]
  {10.1088/0004-637X/703/2/2113}, \href
  {http://adsabs.harvard.edu/abs/2009ApJ...703.2113V} {703, 2113}

\bibitem[\protect\citeauthoryear{{Volonteri} \& {Stark}}{{Volonteri} \&
  {Stark}}{2011}]{Volonteri:2011}
{Volonteri} M.,  {Stark} D.~P.,  2011, \mn@doi [\mnras]
  {10.1111/j.1365-2966.2011.19391.x}, \href
  {http://adsabs.harvard.edu/abs/2011MNRAS.417.2085V} {417, 2085}

\bibitem[\protect\citeauthoryear{{Walter} et~al.,}{{Walter}
  et~al.}{2003}]{Walter:2003}
{Walter} F.,  et~al., 2003, \mn@doi [\nat] {10.1038/nature01821}, \href
  {http://adsabs.harvard.edu/abs/2003Natur.424..406W} {424, 406}

\bibitem[\protect\citeauthoryear{{Wang} et~al.,}{{Wang}
  et~al.}{2010}]{Wang:2010}
{Wang} R.,  et~al., 2010, \mn@doi [\apj] {10.1088/0004-637X/714/1/699}, \href
  {http://adsabs.harvard.edu/abs/2010ApJ...714..699W} {714, 699}

\bibitem[\protect\citeauthoryear{{Wang}, {Wagg}, {Carilli}, {Neri}, {Walter},
  {Omont}, {Riechers}  \& {Bertoldi}}{{Wang} et~al.}{2011}]{Wang:2011}
{Wang} R.,  {Wagg} J.,  {Carilli} C.~L.,  {Neri} R.,  {Walter} F.,  {Omont} A.,
   {Riechers} D.~A.,   {Bertoldi} F.,  2011, \mn@doi [\aj]
  {10.1088/0004-6256/142/4/101}, \href
  {http://adsabs.harvard.edu/abs/2011AJ....142..101W} {142, 101}

\bibitem[\protect\citeauthoryear{{Wang}, {Wagg}, {Carilli}, {Walter},
  {Lentati}, {Fan}, {Riechers}  \& {Bertoldi}}{{Wang} et~al.}{2013}]{Wang:2013}
{Wang} R.,  {Wagg} J.,  {Carilli} C.~L.,  {Walter} F.,  {Lentati} L.,  {Fan}
  X.,  {Riechers} D.~A.,   {Bertoldi} F.,  2013, \mn@doi [\apj]
  {10.1088/0004-637X/773/1/44}, \href
  {http://adsabs.harvard.edu/abs/2013ApJ...773...44W} {773, 44}

\bibitem[\protect\citeauthoryear{{Wei{\ss}}, {Downes}, {Neri}, {Walter},
  {Henkel}, {Wilner}, {Wagg}  \& {Wiklind}}{{Wei{\ss}}
  et~al.}{2007}]{Weis:2007}
{Wei{\ss}} A.,  {Downes} D.,  {Neri} R.,  {Walter} F.,  {Henkel} C.,  {Wilner}
  D.~J.,  {Wagg} J.,   {Wiklind} T.,  2007, \mn@doi [\aap]
  {10.1051/0004-6361:20066117}, \href
  {http://adsabs.harvard.edu/abs/2007A%26A...467..955W} {467, 955}

\bibitem[\protect\citeauthoryear{{Willott}, {McLure}  \& {Jarvis}}{{Willott}
  et~al.}{2003}]{Willott:2003}
{Willott} C.~J.,  {McLure} R.~J.,   {Jarvis} M.~J.,  2003, \mn@doi [\apjl]
  {10.1086/375126}, \href {http://adsabs.harvard.edu/abs/2003ApJ...587L..15W}
  {587, L15}

\bibitem[\protect\citeauthoryear{{Willott} et~al.,}{{Willott}
  et~al.}{2010}]{Willott:2010}
{Willott} C.~J.,  et~al., 2010, \mn@doi [\aj] {10.1088/0004-6256/140/2/546},
  \href {http://adsabs.harvard.edu/abs/2010AJ....140..546W} {140, 546}

\bibitem[\protect\citeauthoryear{{Wu} et~al.,}{{Wu} et~al.}{2015}]{Wu:2015}
{Wu} X.-B.,  et~al., 2015, \mn@doi [\nat] {10.1038/nature14241}, \href
  {http://adsabs.harvard.edu/abs/2015Natur.518..512W} {518, 512}

\bibitem[\protect\citeauthoryear{{da Cunha} et~al.,}{{da Cunha}
  et~al.}{2013}]{da-Cunha:2013}
{da Cunha} E.,  et~al., 2013, \mn@doi [\apj] {10.1088/0004-637X/766/1/13},
  \href {http://adsabs.harvard.edu/abs/2013ApJ...766...13D} {766, 13}

\bibitem[\protect\citeauthoryear{{van der Werf} et~al.,}{{van der Werf}
  et~al.}{2010}]{vanderwerf:2010}
{van der Werf} P.~P.,  et~al., 2010, \mn@doi [\aap]
  {10.1051/0004-6361/201014682}, \href
  {http://adsabs.harvard.edu/abs/2010A%26A...518L..42V} {518, L42}

\makeatother
\end{thebibliography}




\appendix

\section{Visibility analysis}
\label{app:uv}
In Section \ref{sec:continuum.emission} we presented the analysis of the continuum emission. Here we report more details of the fit procedure.

To perform the fits we follow the quick start example reported in the documentation of \texttt{GALARIO}\footnote{\href{https://mtazzari.github.io/galario/quickstart.html}{https://mtazzari.github.io/galario/quickstart.html}}:
\begin{enumerate}
    \item we extract the channel-averaged continuum visibilities from the ALMA Measurement set using the function \texttt{export\_uvtable} included in the the publicly available \texttt{uvplot} package\footnote{\href{https://github.com/mtazzari/uvplot}{https://github.com/mtazzari/uvplot}}.
    \item we define our brightness model as the Sersic profile:
    \begin{equation}
    I(R)=I_e\exp\left\{-b_n\left[\left(\frac{R}{R_{e}}\right)^{(1/n)}-1\right]\right\}
    \end{equation}
    where $R_e$ is the effective (half-light) radius, $I_e$ is the surface brightness at $R=R_e$, and $b_n$ is such that $\Gamma(2n)=2\gamma(2n,b_n)$ and we fix the index $n=1$ (i.e. exponential profile). We compute the brightness profile $I(R)$ on a radial grid ranging from $0.0001''$ to $10''$, with spacing $0.001''$.  
    \item we define the parameter ranges explored by the 40 walkers in the MCMC ensemble sampler. We use log-uniform priors for $I_e$ and $R_e$, $p(\log I_e)=U(-5, 1)$, $p(\log R_e)=U(-2, 0)$ and uniform priors for the other free parameters, $p(i)=U(0, 90)^\circ$, $p(\mathrm{P.A.})=U(180, 360)^\circ$. 
\end{enumerate}

\begin{figure}
    \centering
	\includegraphics[width=0.8\columnwidth]{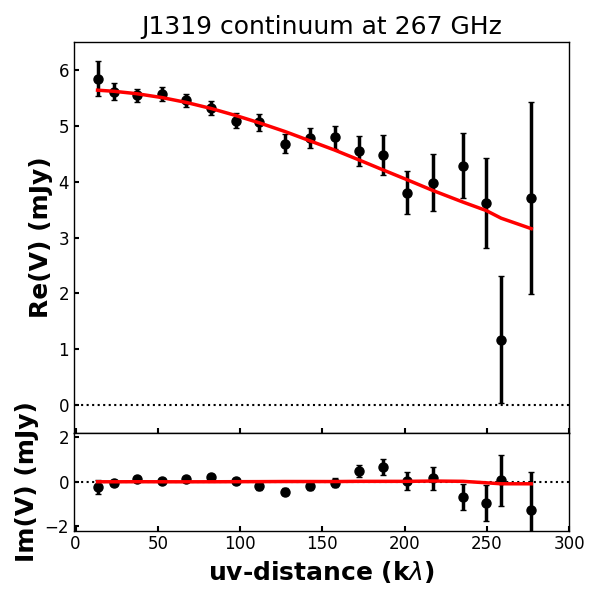}
	\caption{Comparison between the model and the observed visibilities (real and imaginary part) of the J1319 continuum emission at 267\,GHz as a function of deprojected baseline (uv-distance). For the deprojection we used the inferred inclination and P.A. The black dots are the observations, the red line shows the best-fit model.}

\label{fig:app.uv.J1319}
\end{figure}

After the MCMC chain has converged, we assess the goodness of the fit by comparing the best-fitting model and the observations, directly in the plane of the measurements. An immediate way of checking whether a model fits the interferometric data is to produce a so called uv-plot, namely the azimuthal average of the visibilities as a function of deprojected baseline (uv-distance). Figure \ref{fig:uv}
 in Section~\ref{sec:continuum.emission} and ~\ref{fig:app.uv.J1319} here show the uvplot comparing the best-fitting models and the observed visibilities of J2310 and J1319. 
Although the complex visibilities are defined on the uv-plane $V(u, v)$, it is often convenient to represent them in bins of deprojected uv-distance $\rho=\sqrt{u^2+v^2}$. 
It is worth noting that the fit performed with \texttt{GALARIO} allowed us to fit \textit{each single} observed $(u_j, v_j)$ visibility point by computing the corresponding $V_\mathrm{mod}(u_j, v_j)$ of a given model image; the azimuthally averaged view of the bestfit model and of the data given in the uv-plot serves as a benchmark of the goodness of the fit and is inevitably showing less data than it was actually used for the fit. 

To produce the uv-plot we used the \texttt{uvplot} package following the instructions in the \textit{GALARIO} quick start example (see link in the footnotes). To deproject the visibilities we assume inclination and position angle inferred from the fit. 

\begin{figure}
    \centering

   \includegraphics[width=1\columnwidth]{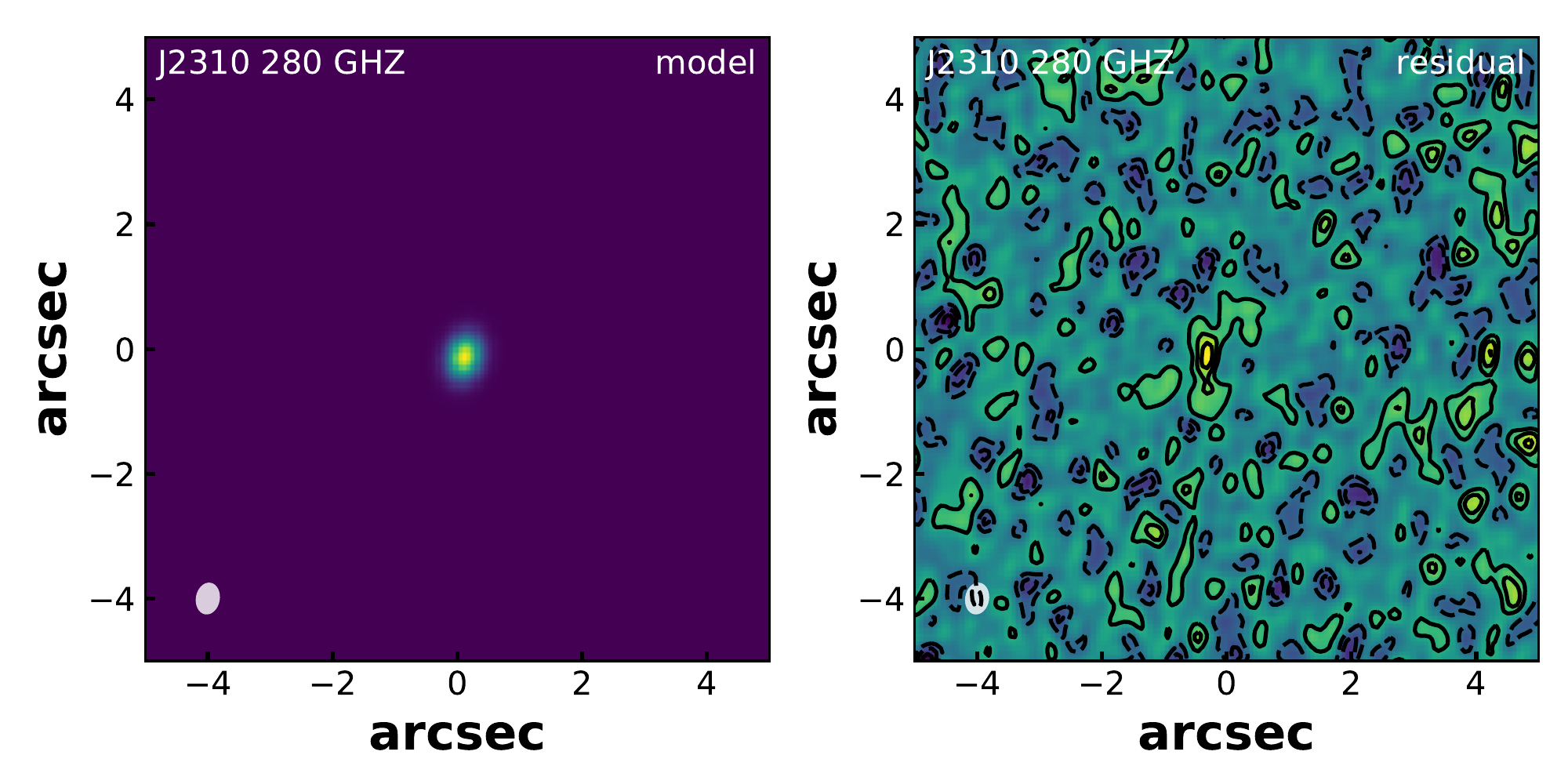}
	\caption{Synthesised image of the model (left) and residual (right) visibilities. Contours in the right panels are in steps of 1(-1)$\sigma$ starting from $\sigma=42 \mu\mathrm{Jy}\,\mathrm{beam}^{-1}$, which is the rms noise.}

\label{fig:app.residuals.J2310}
\end{figure}

\begin{figure}
    \centering
   \includegraphics[width=1\columnwidth]{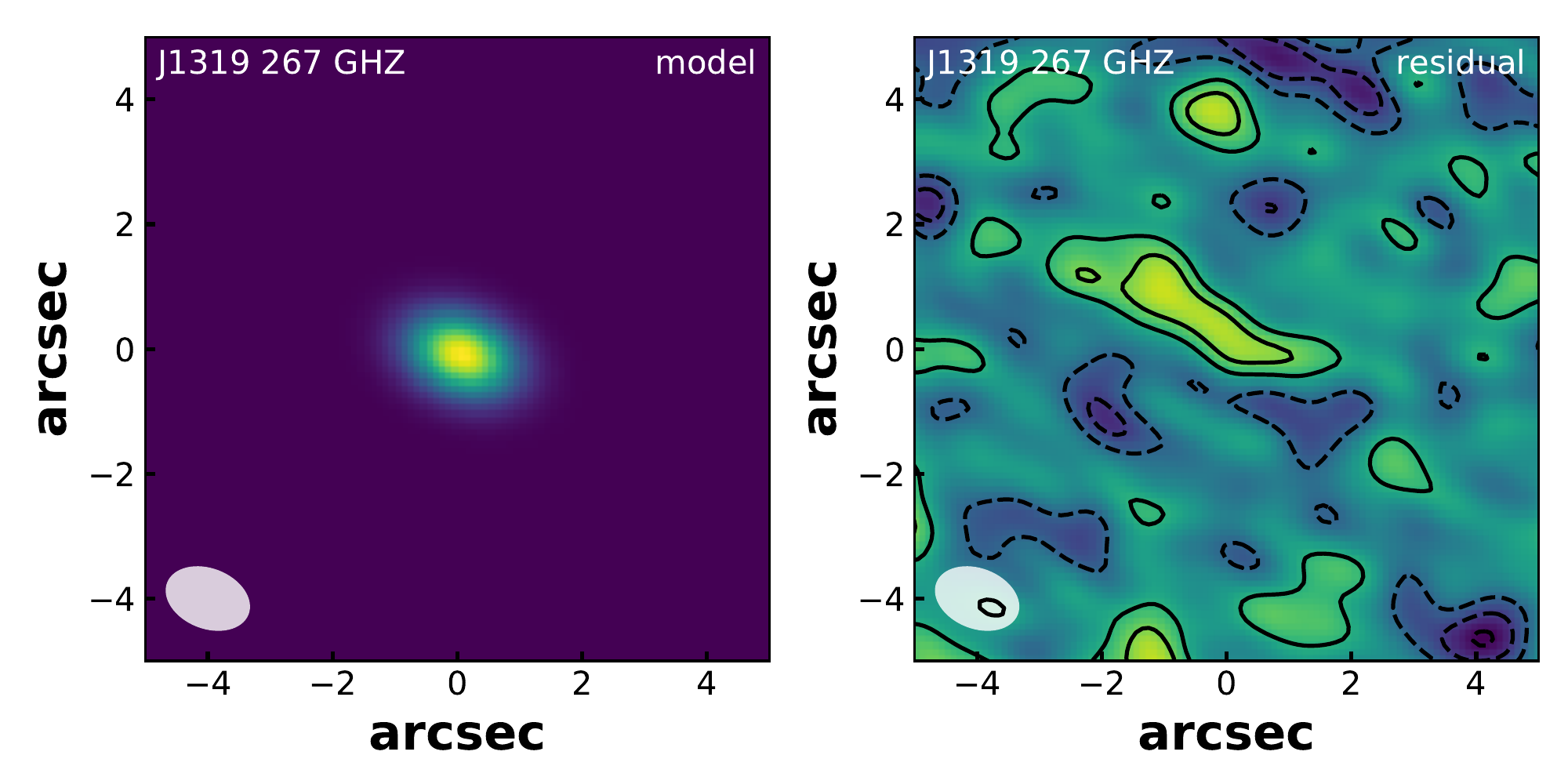}
	\caption{Synthesised image of the residual visibilities. Contours in the right panels are in steps of 1(-1)$\sigma$ starting from $\sigma=43 \mu\mathrm{Jy}\,\mathrm{beam}^{-1}$, which is the rms noise.}

\label{fig:app.residuals.J1319}
\end{figure}

Another way to assess the goodness of the fit is to produce synthesised images of the residual visibilities, namely to compute $V_\mathrm{res}(u_j, v_j)=V_\mathrm{obs}(u_j, v_j)-V_\mathrm{mod}(u_j, v_j)$ for the bestfit model and then obtaining the CLEANed image corresponding to $V_\mathrm{res}$. Figures~\ref{fig:app.residuals.J2310} and \ref{fig:app.residuals.J1319} show the synthesised images of the model and of the residual visibilities. In both cases, the extremely low levels of the residuals (within the 3$\sigma$ noise levels) indicate that the models closely match the measurements.

\section{Optically-thin assumption for dust SED model}
\label{app:thin}

In the optically-thin assumption ($\tau_\nu<<1$) equation~\ref{eq:dust} can be rewritten as: 
\begin{equation}
\label{eq:thin}
\begin{split}
 S^{\rm obs}_{\nu/(1+z)} & \simeq \frac{\Omega}{(1+z)^3}\left[B_{\nu}(T_{\rm dust}(z))-B_{\nu}(T_{\rm CMB}(z)\right]\tau_\nu \\
& = \frac{(1+z) A_{\rm galaxy}}{D_{\rm L}^2}\left[B_{\nu}(T_{\rm dust}(z))-B_{\nu}(T_{\rm CMB}(z)\right]\frac{M_{\rm dust}}{A_{\rm galaxy}}k_\nu\\
& = \frac{(1+z)M_{\rm dust}}{D_{\rm L}^2}\left[B_{\nu}(T_{\rm dust}(z))-B_{\nu}(T_{\rm CMB}(z)\right]k_\nu
\end{split}
\end{equation}

Similarly to the analysis performed in Section~\ref{sec:fir} we 
assume $k_\nu = 0.45(\nu/250{\rm GHz)^\beta \ cm^{2} \ g^{-1}}$ and fit the continuum measurements of the three AGN host galaxies. The best-fitting results (Table~\ref{tab:thin}1), which have been obtained from a MCMC analysis, are in agreement with those from previous studies \citep{Wang:2013, Leipski:2013, Shao:2019}. 

By comparing the results obtained from equation~\ref{eq:dust}, in which dust opacity is accounted for, and those from the optical thin model (equation~\ref{eq:thin}), we note that while the dust mass and emissivity index measurements are consistent within 3$\sigma$, the best-fitting dust temperature values strongly depend on the dust opacity assumptions. 

\begin{table}
\label{tab:thin}
	\centering
	\caption{Results of the SED fitting (assuming $k_0=0.45~\rm cm^{2}~g^{-1}$,$\nu_0=250~\rm GHz$, and optical thin modified black body profile (equation~\ref{eq:thin})}
	\begin{tabular}{lccc } 
		\hline
				 & J2310 & J1319 & J1148 \\
		\hline
		\hline
		\multicolumn{4}{c}{Dust parameters (optical thin assumption)}\\
		\hline
		\hline
		Log(M$_{\rm dust}$/M$_\odot$) & 9.08$\pm0.06$ & 8.9$\pm0.2$  & 8.6$\pm0.1$\\
		T$_{\rm dust}$ [K] & $41^{+5}_{-3}$  & $55\pm15$  & $63^{+12}_{-10}$\\
		$\beta$ & $1.7\pm0.1$ & $1.3^{+0.3}_{-0.2}$ & $1.4\pm0.3$\\
		L$_{\rm FIR}$ [10$^{13}$ \lsun] & $1.6^{+0.4}_{-0.3}$ & $1.7^{+1.6}_{-0.7}$  & $2.3^{+0.8}_{-0.6}$\\
			\hline

	\end{tabular}
\end{table}

\section{Local CO SLEDs}
\label{app:local_co_sled}
Table~\ref{tab:local_co_sled} shows the list of local AGN and starbursts galaxies that have been adopted to obtain the average local CO SLEDs. The intensity of CO lines of each galaxy are reported by \cite{Rosenberg:2015}, \cite{Mashian:2015}, and \cite{Mingozzi:2018}.

\begin{table}
	\centering
	\caption{List of local AGN and starburst galaxies having CO multi-line observations. }
	\label{tab:local_co_sled}
	\begin{tabular}{|c c |} 
		\hline
				AGN & Starburst galaxies \\
		\hline
			NGC 4945 & NGC 253\\
			Circinus & M83  \\
			Mrk 231  & M82\\
			NGC 6240 & IC 694\\
			NGC 3690 & NGC 4418\\
			NGC 1068 & MCG+12-02-001\\
			NGC 34 & NGC 1614\\
			IC 1623 & NGC 2146\\
			NGC 1365 & NGC 3256 \\
			IRASF 05189-2524 & ESO 320-G030\\
			NGC 2623 & ESO 173-G015\\
			Arp 229 & IRASF 17207-0014\\
			IRAS 13120-5453 & IC 4687\\
			NGC 5135 & NGC 7552\\
			Mrk 273 & NGC 7771\\
			NGC 6240 & Mrk 331\\
			NGC 7469 &  \\
					\hline

  	\end{tabular}
    
     References: \cite{Rosenberg:2015}, \cite{Mashian:2015}, and \cite{Mingozzi:2018}.
\end{table}


\bsp	
\label{lastpage}
\end{document}